\documentclass[11pt]{article}
\usepackage{setspace}
\usepackage[top=1in, bottom=1in, left=1in, right=1in]{geometry}
\usepackage{amsmath,amsfonts,amssymb,bm}
\usepackage{amsthm}
\usepackage{xcolor}
\usepackage{environ}
\usepackage{comment}
\usepackage{booktabs}
\usepackage{float}
\usepackage{graphicx}
\usepackage{url}
\usepackage{authblk}  
\usepackage{natbib}
\usepackage{bibunits}

\usepackage[colorlinks,citecolor=blue,linkcolor=blue,urlcolor=blue]{hyperref}
\usepackage{cleveref}

\def\eqref#1{equation~\ref{#1}}

\def\1{\bm{1}}
\newcommand{\train}{\mathcal{D}}

\newcommand{\balpha}{\boldsymbol{\alpha}}

\newcommand{\vzero}{{\vec{0}}}

\newcommand{\diff}{\ensuremath{\mathrm{d}}}

\def\vzero{{\bm{0}}}

\def\vmu{{\bm{\mu}}}

\def\vw{{\bm{w}}}
\def\vx{{\bm{x}}}
\def\vy{{\bm{y}}}

\def\vF{{\bm{F}}}
\def\vG{{\bm{G}}}
\def\vGh{{\bm{B}}}
\def\vH{{\bm{H}}}

\def\vC{{\bm{C}}}

\def\vxh{{\hat{\bm{x}}}}
\def\vyh{{\hat{\bm{y}}}}

\def\vwh{{\hat{\bm{w}}}}

\def \vmuh{\hat{\bm{\mu}}}
\def\evxh{{\hat{\evx}}}
\def\evyh{{\hat{\evy}}}
\def\evmuh{{\hat{\evmu}}}
\def\vdotxh{{\hat{\dot{\bm{x}}}}}
\def\vdotyh{{\hat{\dot{\bm{y}}}}}

\def\vdotmuh{{\hat{\dot{\bm{\mu}}}}}
\def\Mx{{M}}
\def\My{{P}}

\def\evmu{{\mu}}

\def\evx{{x}}
\def\evy{{y}}
\def\evz{{z}}

\def\mF{{\bm{F}}}

\def\mH{{\bm{H}}}

\DeclareMathAlphabet{\mathsfit}{\encodingdefault}{\sfdefault}{m}{sl}
\SetMathAlphabet{\mathsfit}{bold}{\encodingdefault}{\sfdefault}{bx}{n}

\def\sF{{\mathbb{F}}}
\def\sG{{\mathbb{G}}}

\def\sS{{\mathbb{S}}}

\newcommand{\E}{\mathbb{E}}

\newcommand{\rademacher}{\mathcal{R}}
\newcommand{\R}{\mathbb{R}}

\newcommand{\simiid}{\overset{\mathrm{i.i.d.}}{\sim}}

\DeclareMathOperator{\tint}{{\textstyle \int}}
\DeclareMathOperator{\tsum}{{\textstyle \sum}}

\newcommand{\rkhs}{\mathcal{H}}
\newcommand{\sobolev}{\mathcal{W}}

\newif\ifcomments
\commentstrue
\ifcomments\newcommand{\comments}[1]{#1}\else\newcommand{\comments}[1]{}\fi
\definecolor{clrgp}{rgb}{.9,0,.9}

\newif\ifrestating
\restatingfalse
\NewEnviron{restatable}[3]{%
  \expandafter\xdef\csname restatethis@#2\endcsname{%
    \unexpanded\expandafter{\BODY}%
  }%
  \begin{#1}[#3]
    \label{#2}
    \BODY
  \end{#1}%
  \newtheorem*{#2}{\Cref{#2} (Restated)}%
}
\newcommand{\restate}[1]{%
  \restatingtrue
  \begin{#1}\csname restatethis@#1\endcsname\end{#1}%
  \restatingfalse
}

\theoremstyle{plain}

\newtheorem{theorem}{Theorem}

\newtheorem{lemma}[theorem]{Lemma}

\newtheorem{corollary}[theorem]{Corollary}

\newtheorem{assumption}{Assumption}

\theoremstyle{definition}
\newtheorem{definition}{Definition}

\crefname{theorem}{Theorem}{Theorems}
\crefname{section}{Section}{Sections}
\crefname{assumption}{Assumption}{Assumptions}
\crefname{figure}{Figure}{Figures}
\crefname{definition}{Definition}{Definitions}
\crefname{lemma}{Lemma}{Lemmas}
\crefname{proposition}{Proposition}{Propositions}
\crefname{equation}{Equation}{Equations}
\crefname{table}{Table}{Tables}

\title{Solving Models of Economic Dynamics with Ridgeless Kernel Regressions}

\author[1]{Mahdi Ebrahimi Kahou}
\author[2]{Jesse Perla}
\author[3,4]{Geoff Pleiss}
\affil[1]{Department of Economics, Bowdoin College.}
\affil[2]{Vancouver School of Economics, University of British Columbia.}
\affil[3]{Department of Statistics, University of British Columbia.}
\affil[4]{Vector Institute.}

\date{\today} 

\begin{document}
	\maketitle
	\renewcommand{\thefootnote}{} 
	\footnotetext{We thank Misha Belkin, Amy Greenwald, William Jungerman, Lilia Maliar, Christian Matthes, Hikaru Saijo, Felipe Schwartzman, and James Yu. We are especially grateful to Rachel Childers for comments and discussions. We gratefully acknowledge support from a SSHRC Insight Grant number 435-2023-0119. Replication materials: \url{https://github.com/HighDimensionalEconLab/kernel_econ_alignment}.}
	\renewcommand{\thefootnote}{\arabic{footnote}} 
	\begin{abstract}
		This paper proposes a ridgeless kernel method for solving infinite-horizon, deterministic, continuous-time models in economic dynamics, formulated as systems of differential-algebraic equations with asymptotic boundary conditions (e.g., transversality).  Traditional shooting methods enforce the asymptotic boundary conditions by targeting a known steady state---which is numerically unstable, hard to tune, and unable to address cases with steady-state multiplicity.
		Instead, our approach solves the underdetermined problem without imposing the asymptotic boundary condition, using regularization to select the unique solution fulfilling transversality among admissible trajectories.  In particular, ridgeless kernel methods recover this path by selecting the minimum norm solution, coinciding with the non-explosive trajectory.
		We provide theoretical guarantees showing that kernel solutions satisfy asymptotic boundary conditions without imposing them directly, and we establish a consistency result ensuring convergence within the solution concept of differential-algebraic equations. Finally, we illustrate the method in canonical models and demonstrate its ability to handle problems with multiple steady states.
	\end{abstract}
	
	\vspace{0.5em}
	

	\section{Introduction}
	\label{sec:intro}
	This paper proposes using ridgeless kernel regression to solve a broad class of infinite-horizon, deterministic, continuous-time models in economic dynamics.
The main computational challenge in these models is satisfying asymptotic boundary conditions, typically arising from transversality conditions in the embedded optimal control problem.
Unlike shooting algorithms that aim toward a known steady state, we show that kernel methods can satisfy these asymptotic conditions by selecting the least explosive trajectory among all candidate solutions---and without even calculating the steady-state, let alone imposing it as a condition.
This yields robust and computationally efficient algorithms, even under steady-state multiplicity and hysteresis, more stable than parametric approaches such as deep learning, and often easier to tune than classic shooting methods.

Beyond capturing forward-looking behavior, asymptotic boundary conditions are essential for well-posedness in the sense of Hadamard \citep{hadamard1902problemes,hadamard1932probleme}---ensuring existence, uniqueness, and continuous dependence on initial conditions.
Without them, the system may admit a continuum of trajectories satisfying both the initial conditions and the differential-algebraic equations (DAEs), thereby violating well-posedness.

The central idea of this paper is to treat the problem as underdetermined: impose initial conditions and laws of motion, but not the asymptotic boundary conditions, and then use regularization to select among the resulting trajectories.\footnote{See \citet{tikhonov1963solution} and \citet{doi:10.1137/1021044} for classic treatments of regularization in ill-posed problems.
Our approach is also related to implicit and explicit regularization in deep learning \citep{spooky}.}
For a large class of models, the unique valid solution is also the only non-explosive one; all others fail transversality.
By choosing an appropriate function norm and penalizing it, kernel methods recover this non-explosive solution---thereby satisfying the asymptotic boundary conditions.

{\it Kernel Methods.~}
Kernel methods are a natural fit for this task since they provide a formal framework for defining and penalizing function norms through the theory of the Reproducing Kernel Hilbert Space (RKHS) induced by the choice of the kernel.  In particular, we focus on ridgeless kernel methods \citep{belkin2019reconciling}, which select the minimum norm solution among all solutions that perfectly satisfy the differential equation at some finite number of points.

A ``kernel machine'' is a non-parametric approximation where a function is represented as a weighted sum of the distances to all existing data (i.e., a kernel).\footnote{See \cite{rasmussen2006gaussian} and \cite{probml1} for more details on kernel methods, Reproducing Kernel Hilbert Spaces, and the Representer Theorems.}  For goals ranging from empirical risk-minimization to solving a system of function equations, kernels and RKHS map an infinite dimensional problem in a function space to a finite dimensional problem in the weights of the kernel machine for all data.

{\it Related Work.~} Our approach is connected with both traditional methods for solving linearized systems and to a recent literature that uses ML to solve nonlinear optimal control problems.

Perturbation solutions to models of economic dynamics use classic stability analysis from linear-quadratic (LQ) control to find solutions that satisfy asymptotic boundary conditions (e.g., \citet{blanchard1980solution}). These methods exploit the linearity of time-invariant policies in LQ control, ensuring stability by ruling out explosive roots inconsistent with transversality and selecting the unique stabilizing solution. Our paper draws inspiration from this broader approach: algorithms that select non-explosive roots lie on the solution manifold and automatically satisfy transversality conditions.

The use of ML methods is becoming increasingly popular for solving and estimating economic models. Applications span a wide range, including wealth inequality \citet{han2022deepham}, financial frictions \citet{https://doi.org/10.3982/ECTA18180}, the heterogeneous impacts of climate change \citet{barnett2023deep}, portfolio choice problems \citet{azinovic2023economics}, heterogeneous agent New Keynesian models \citet{kase2022estimating}, human capital accumulation in the labor market \citet{jungerman2023}, and labor market dynamics in search and matching environments \citet{DeepSAM}. These models often rely on neural networks \citet{AzinovicEtAl, maliar2021deep, NBERw28981, spooky} and even some work with Gaussian Processes \citet{SCHEIDEGGER201968}. 

Recent work in optimal control explores how to achieve stable solutions using ML-based methods \cite{9887885,nakamura2022neural,chen2023deep,NEURIPS2019_2647c1db}. More directly within economics, \cite{spooky} discusses the intuitive connection between the inductive bias of deep neural networks and turnpikes \cite{turnpikes} in dynamic economic models, but does not provide a formal theory or consider kernel methods. Our paper contributes to this literature, providing a formal argument	on why the inductive bias of ML algorithms promotes stability in infinite-horizon control when using particular kernel machines.

{\it Contributions.~}
Core results of our paper include:
\begin{itemize}
	\item {\bf Inductive bias alignment:} theoretical and empirical evidence that the minimum-norm implicit bias of kernel methods aligns with many asymptotic boundary conditions found in economic problems;
	\item {\bf Learning the right set of steady states:} evidence that kernel machines identify the steady states of dynamical systems—the ones corresponding to the optimal solution—which leads to highly accurate generalization outside the training data, even without enforcing asymptotic boundary conditions;
	\item {\bf Consistency of ML estimates:} guarantees that the approximation error of our kernel methods can be bounded, with the method converging to the true minimum norm solution (and thus solutions satisfying asymptotic boundary conditions) as training data increases; and
	\item {\bf Robustness and speed:} demonstrations that kernel machines can be competitive in both speed and robustness with traditional methods for modeling economic systems, even on small-scale problems.
\end{itemize}

{\it Structure.~} The remainder of this paper is structured as follows. \cref{sec:setup} describes the class of economic models and provides assumptions on primitives which lead to explosive solutions violating transversality.  \cref{sec:method} describes how kernel methods map to our class of problems, and discusses cases where a min norm solution is sufficient to fulfill transversality.  \cref{thm:consistency} provides a consistency result showing how kernel methods converge to the min norm solution, aligning with the solution concept of the DAE itself.  Results from our core applications and future directions are presented in \cref{sec:results}, while \cref{sec:Conclusion} concludes the paper.

	\section{Setup}
	\label{sec:setup}
	\label{section-setup}
We focus on an important class of dynamic problems in economics and finance: deterministic, continuous-time systems that arise from discounted infinite-horizon optimal control problems, together with algebraic constraints that encode conditions such as instantaneous market clearing.\footnote{An inherent characteristic of discounted infinite-horizon optimal control problems,
	whether deterministic or stochastic,
	are asymptotic boundary conditions such as transversality conditions.  These conditions can be
	formulated sequentially or recursively in a state space, and in continuous- or discrete-time (see discussions of necessary conditions in \citet{michel1982transversality,BENVENISTE19821,van2007optimal}).}

{\it Problem class.~}
In these models, the first-order necessary conditions of the underlying decision problems can be stacked into a system of ordinary differential and algebraic equations (DAEs). 
 The variables can be partitioned into three vectors: \emph{state variables} $\vx(t)\in\R^{\Mx}$, with an initial condition $\vx_0$; \emph{co-state variables} $\vmu(t)\in\R^{\Mx}$, with accompanying asymptotic boundary conditions that typically arise from the embedded control problem; and \emph{jump variables} $\vy(t)\in\R^{\My}$, which relate the state and co-state variables (e.g., co-state $=$ the marginal utility of consumption) and/or impose intratemporal constraints (e.g., market clearing conditions).\footnote{When present, jump variables constrain the solution manifold and are not matched with boundary or initial values. The connection between the number of \emph{jump variables} and stability local to a steady-state is discussed in \citet{blanchard1980solution}.  While this paper analyzes the convergence for DAEs, the computational methods can be used for systems augmenting inequality constraints in addition to \cref{eq:f3} and differential inclusions in \cref{eq:f1,eq:f2}.}
Canonical examples of such equations arise from applying Pontryagin's Maximum Principle to the present-value Hamiltonian of a dynamic optimization problem, as in \citet{acemoglu2008introduction}. In that case, $\vmu(t)$ represents the present-value Lagrange multipliers associated with the state variables $\vx(t)$. The dynamical system with primitives $\vF, \vG : \R^{\Mx}\times \R^{\Mx}\times\R^{\My} \to \R^{\Mx}$ and $\vH : \R^{\Mx}\times \R^{\Mx}\times\R^{\My} \to \R^{\My}$ with a discount rate $r > 0$ is
\begin{align}
	\dot{\vx}(t) &= \vF(\vx(t),\vmu(t), \vy(t)),\label{eq:f1}\\
	\dot{\vmu}(t) &= r \vmu(t) - \vmu(t) \odot \vG(\vx(t),\vmu(t), \vy(t)),\label{eq:f2}\\
	\mathbf{0} &= \vH(\vx(t), \vmu(t), \vy(t)),\label{eq:f3}\\
	\intertext{subject to $\Mx$ initial values, and $\Mx$ asymptotic boundary conditions (i.e., transversality conditions)}
	\vx(0) &= \vx_0 \label{eq:f-iv}\\
	\mathbf{0} &= \lim_{t\to\infty}e^{-r t} \vx(t) \odot \vmu(t) \label{eq:f4}
\end{align}
The symbol $\odot$ denotes element-wise multiplication between vectors of the same dimension.

This system is an autonomous, semi-explicit differential-algebraic equation (DAE) with a mix of initial conditions, \cref{eq:f-iv}, and asymptotic boundary conditions, \cref{eq:f4}.

{\it Example: Neoclassical Growth Model.~}
The canonical example of this class in macroeconomics is the neoclassical growth model.  Fitting to our notation define: capital $x(t)$, consumption $y(t)$, flow utility $\log(y)$, present-value co-state variable $\mu(t)$, discount rate $r > 0$, depreciation rate $0 < \delta < 1$, and a monotonically increasing and strictly concave production function
$f(x)$ where $f(0) = 0$.  Writing the standard equations in our notation with $x, \mu, y \in \mathbb{R}^1$,\footnote{
The derivation follows the standard planning problem maximizes the lifetime discounted utility of consumption:
$\int_0^\infty e^{-rt} u\left(c(t)\right) dt$, with $u(c) = \log(c)$, subject to the law of motion for capital: $\dot{k}(t) = f\left(k(t)\right) - \delta k(t) - c(t)$, where the production function is
$f(k) = k^a$ with $0 < a < 1$, and the depreciation rate satisfies $0 < \delta < 1$.
In this case, the present-value Hamiltonian is
$ u\left(c(t)\right) + \mu(t)\left[f\left(k(t)\right) - \delta k(t) - c(t)\right]$.  The DAE follows from applying Pontryagin's Maximum Principle along with a standard transversality condition, and mapped to our notation (see \citet{acemoglu2008introduction}).
}

\begin{align}
	\dot{x}(t) &= f\left(x(t)\right) - \delta x(t) - y(t):= F(x(t), \mu(t), y(t))\label{eq:ngm-f1} \\
	\dot{\mu}(t) &= r \mu(t) - \mu(t) \underbrace{\left[f'\left(x(t)\right)- \delta \right]}_{:= G(x(t), \mu(t), y(t))}\label{eq:ngm-f2}  \\
	 0 &= \mu(t)y(t) -1:= H(x(t), \mu(t), y(t))\label{eq:ngm-f3}\\
	x(0) &= x_0\label{eq:ngm-iv}\\
	0 &= \lim_{t \to \infty} e^{-rt} \mu(t) x(t)\label{eq:ngm-f4}
\end{align}
Next we will analyze the key conditions required for our algorithm, and demonstrate the intuition with this running example.

\subsection{Assumptions for Unique and Bounded Solutions}
Key necessary conditions for the DAE to be \emph{well-posed} are that for the given $\vx_0$, there exists a unique $\vmu(0)$ fulfilling the asymptotic boundary condition \cref{eq:f4}.  \cref{ass:regularity} provides further assumptions to ensure that there exists a unique $\vy(0)$ fulfilling \cref{eq:f3} given $\vx(0)$ and $\vmu(0)$.

\begin{assumption}[Conditions for Well-posedness and Regularity]\label{ass:regularity}
	Assume that 
	\begin{itemize}
		\item $\vF$, $\vG$, and $\vH$ in \cref{eq:f1,eq:f2,eq:f3,eq:f-iv}
		are Lipschitz with respect to $\Vert \cdot \Vert_\infty$;
		\item $\vF$ and $\vG$ have Lipschitz first derivatives and $\vH$ has Lipschitz first and second derivatives;
        \item The Jacobian of $\vH$ with respect to $\vy$ is nonsingular along the relevant trajectories:
        $\det\!\big(\nabla_{\vy}\vH(\vx,\vmu,\vy)\big)\neq 0$,
		and its inverse is Lipschitz continuous with a Lipschitz first derivative in the domain $t \in [0, T]$.
	\end{itemize}
\end{assumption}

 By the implicit function theorem the last condition in \cref{ass:regularity} implies a unique, locally Lipschitz, map $\vy=\vy(\vx,\vmu)$ fulfilling $\vH(\vx, \vmu, \vy)=\mathbf{0}$---so no initial/boundary condition for $\vy(0)$ is required.  This ensures that the system is a semi-explicit DAE with index~$1$.\footnote{Algorithms that reduce the index of a semi-explicit DAE to an ODE, such as the Pantelides algorithm, may augment $\vx(t)$ and $\vmu(t)$ to eliminate the algebraic equation (cf. \cref{eq:f3}). While this procedure is often carried out by hand in macroeconomics, we will instead work directly with the more natural DAE formulation.}

The conditions in \cref{ass:regularity} ensure uniqueness given an initial condition---eliminating sources of multiplicity that may arise due to the jump variables, $\vy(t)$.\footnote{For example, some models have multiple $\left(\vx(t),\vmu(t),\vy(t)\right)$ fulfill \cref{eq:f1,eq:f2,eq:f3,eq:f4,eq:f-iv} for a given $\vx_0$.  These often come out of coordination failures in monetary economics and multiplicity in rational expectations equilibria.  While these models with multiplicity are not considered in our paper, there may be hysteresis and multiplicity of steady-states---i.e., $\lim_{t\to\infty} \vx(t)$ may depend on $\vx_0$.  For example, these methods could solve transition paths and steady-states in dynamic models of trade, such as \cite{RAVIKUMAR201993}, where steady-state depends on the current account initial conditions.}

In addition, we add another standard assumptions on problem formulation to ensure that both the state and co-state variables in a solution are bounded and strictly positive.\footnote{See \citet[p.~51]{arrow1970public} for when these assumptions hold.  In practice, it is usually sufficient to use a present-value rather than current-value Hamiltonian, and solve a de-trended model in cases with growth. 
}

\begin{assumption}[Bounded Solutions]\label{ass:interior}
	Assume that:
	\begin{itemize}
		\item For any given $\vx_0$, there is a unique solution, $\vx(t), \vmu(t), \vy(t)$ to \cref{eq:f1,eq:f2,eq:f3,eq:f-iv,eq:f4}.
		\item There exist bounds such that, $\mathbf{0} < \underline{\vx} < \vx(t) < \bar{\vx} < \infty, \mathbf{0} < \underline{\vmu} < \vmu(t)< \bar{\vmu} < \infty,$ and $0 < \underline{\vy}< \vy(t) < \bar{\vy} < \infty$ for all $t > 0$ on the solution path for a given $\vx_0$.
	\end{itemize}

\end{assumption}

\subsection{Shooting Methods and Explosive Non-Solutions}\label{sec:explosive-solutions}
Before we explore the kernel-based methods in \cref{sec:method} and solve the neoclassical growth model in \cref{sec:neoclassical-growth-model}, we will analyze shooting methods to provide intuition on our solution algorithm.

The central computational challenge is that applying the boundary-value \cref{eq:f4} numerically is not directly feasible, as it is asymptotic.  Shooting methods, and related algorithms solving for an implicit equation with a finite-horizon BVP Methods, use the insight that if we could replace the asymptotic boundary condition in \cref{eq:f4} with the right $\vmu(0) = \vmu_0$, then \cref{ass:interior,ass:regularity} ensure the solution is a well-posed initial value problem (IVP)---routinely solved for DAEs with millions of equations using software such as \cite{hindmarsh2005sundials}.

First, we need to contrast two types of functions fulfilling the ODEs:  a \textit{solution} trajectory fulfills the full set of well-posed equations \cref{eq:f1,eq:f2,eq:f3,eq:f-iv,eq:f4}; and a \textit{non-solution} trajectory fulfills the ill-posed system \cref{eq:f1,eq:f2,eq:f3,eq:f-iv}, but fails the asymptotic boundary condition \cref{eq:f4}.  \cref{ass:regularity,ass:interior} ensure that we have a unique bounded solution, but there are usually a continuum of non-solutions which can each be associated with a different $\vmu(0)$. The solution is indexed by its initial condition $\vmu_0$, a single point inside an $\Mx$-dimensional set of non-solutions.

{\it Shooting methods.~}
In many applications it is straightforward to compute a steady state $\vx(\infty),\, \vmu(\infty),\, \vy(\infty)$, which is bounded by \cref{ass:interior}. A shooting algorithm takes an initial co-state $\tilde{\vmu}_0$, integrates the dynamics to a large $T$ using a standard DAE/ODE IVP solver, and evaluates
$$
\Psi(\tilde{\vmu}_0;T) \equiv
\begin{bmatrix}
\vx(T;\tilde{\vmu}_0) - \vx(\infty)\\
\vmu(T;\tilde{\vmu}_0) - \vmu(\infty)\\
\vy(T;\tilde{\vmu}_0) - \vy(\infty)
\end{bmatrix}^{\top}.
$$
One then finds $\tilde{\vmu}_0$ such that $\Psi(\tilde{\vmu}_0;T) \approx \mathbf{0}$ using a root–finding method (e.g., bisection, Newton). The transversality condition in \cref{eq:f4} is implicitly used in calculating the steady-state, but it is always present. The stability of this procedure is governed by the Jacobian $\nabla\Psi(\tilde{\vmu}_0;T)$.\footnote{Related approaches convert the DAE into a finite-horizon boundary-value problem (BVP) by discretizing time (e.g., via finite differences) and solving a nonlinear system that enforces the initial conditions, artificial terminal conditions, and the dynamics. Although often preferable to shooting, the resulting Jacobian has similar conditioning because the system effectively embeds the same $\Psi(\cdot;T)$ structure.}

{\it Challenges with shooting methods.~}
Even when the steady state is easily calculated, in practice the algorithm is unstable and highly sensitive to both the initial guess $\tilde{\vmu}_0$ and the chosen horizon $T$.\footnote{In many important applications in growth, trade, international economics, and spatial economics, calculating the steady state itself is difficult. Moreover, there may be initial-condition dependence and multiple steady states. In such cases, shooting methods require first solving for all candidate steady states $x(\infty)$ given $x_0$, analyzing fixed-point stability via the Hessians of \cref{eq:f1,eq:f2,eq:f3}, and then partitioning $\R^{\Mx}$ into basins of attraction. This process is infeasible outside of the simplest, low-dimensional settings.}

If $T$ is too small, the algorithm may converge but produces a biased approximation because it forces trajectories to approach the steady state too quickly. If $T$ is too large, then non-solution trajectories—those failing \cref{eq:f4}—separate rapidly from the true solution. In that case the Jacobian $\nabla\Psi(\tilde{\vmu}_0;T)$ becomes ill-conditioned: $\Psi(\tilde{\vmu}_0;T)$ is close to zero in a small neighborhood around the true $\vmu_0$ and diverges sharply outside it. This lack of smoothness renders root-finding highly sensitive in higher dimensions. Thus one faces a tradeoff: a small $T$ yields stability but bias, while a large $T$ is unbiased but numerically unstable.

This sensitivity is inherent to optimal control problems due to their saddle-path structure: all trajectories except those along the unique solution manifold diverge. For this problem class the situation is even sharper: \cref{thm:divergence-speed} shows that under our assumptions, any non-solution trajectory diverges at least as fast as the exponential discounting rate $r$ appearing in the transversality condition \cref{eq:f4}. Consequently, even small deviations in the initial condition $\tilde{\vmu}_0$ accumulate exponentially over time, directly leading to instability in computing $\nabla \Psi(\cdot;T)$.\footnote{Formally, the condition number of $\nabla \Psi(\cdot;T)$, the ratio of the largest to smallest singular values, will diverge with $T$. It is worth noting that if the focus is only on transition dynamics, the explicit calculation of the steady state is not strictly required beyond providing a target for the shooting method. While $\Psi(\cdot;T)$ uses the steady state for guidance, its main role is to steer the algorithm away from explosive non-solutions—and it can be discarded once a valid solution is found. We will exploit this observation in our kernel-based methods.}


\begin{restatable}{theorem}{thm:divergence-speed}{}[Divergence Rate of Non-Solutions]
Let $\tilde{\vmu}_0$ be the initial condition associated with a non-solution, and
let $\vy(\vx,\tilde{\vmu})$ denote the solution of \cref{eq:f3} given \cref{ass:regularity}, i.e.,
$
\vH\!\left(\vx, \tilde{\vmu}, \vy(\vx,\tilde{\vmu})\right) = \mathbf{0}.
$
Suppose there exist points $\tilde{\vx}^* \in \R^{M}$ and $\tilde{\vy}^* \in \R^{P}$ such that
$$
\lim_{\tilde{\vmu} \to \infty} \vy(\tilde{\vx}^*,\tilde{\vmu}) = \tilde{\vy}^*, \quad
\lim_{\tilde{\vmu} \to \infty} \vF\!\left(\tilde{\vx}^*, \tilde{\vmu}, \vy(\tilde{\vx}^*,\tilde{\vmu})\right) = \mathbf{0}, \quad
\lim_{\tilde{\vmu} \to \infty} \vG\!\left(\tilde{\vx}^*, \tilde{\vmu}, \vy(\tilde{\vx}^*,\tilde{\vmu})\right) \leq \mathbf{0}.
$$
Then
$$
\lim_{t \to \infty}
\frac{\dot{\tilde{\vmu}}^{(m)}(t)}{\tilde{\vmu}^{(m)}(t)} \;\geq\; r,
\qquad \text{ for some } m = 1,\ldots,M.
$$
Furthermore, if $~\lim_{\vmu \to \infty} \vG\!\left(\tilde{\vx}^*, \vmu, \vy(\tilde{\vx}^*,\vmu)\right)^{(m)} < 0$ some $m$, then
$$
\lim_{t \to \infty}
\frac{\dot{\tilde{\vmu}}^{(m)}(t)}{\tilde{\vmu}^{(m)}(t)} \;> \; r,
$$
\end{restatable}

\begin{proof}
	Rewrite \cref{eq:f2} componentwise and take limits
	\begin{align*}
	\lim_{t \to \infty}\frac{\dot{\vmu}^{(m)}(t)}{\vmu^{(m)}(t)} &= \lim_{t \to \infty}  \left(r - \vG(\vx(t), \vmu(t), \vy(t))^{(m)}\right) = \lim_{\vmu \to \infty} r - \vG(\tilde{\vx}^*, \vmu, \tilde{\vy}^*)^{(m)}.
	\end{align*}
	Note that if $\lim_{\vmu \to \infty}\vG(\tilde{\vx}^*, \vmu, \tilde{\vy}^*)^{(m)} < 0$ for some $m$, then $\lim_{t \to \infty} \frac{\dot{\tilde{\vmu}}^{(m)}(t)}{\tilde{\vmu}^{(m)}(t)}  > r$
	
\end{proof}

This result confirms the intuition for why shooting methods are difficult to use in practice. Not only do the non-solutions diverge, they do so exponentially, at least as fast as the transversality condition in \cref{eq:f4}. Consequently, a small error in the initial condition at time zero, $\tilde{\vmu}_0$, accumulates exponentially over time---which leads to the numerical instability of calculating $\nabla \Psi(\cdot;T)$ in shooting methods.
.

{\it Verification for neoclassical growth.}
The condition in \cref{thm:divergence-speed} is easily verified in the neoclassical growth model. If $\tilde{\mu}(t)\to\infty$, then by \cref{eq:ngm-f3} we have $\tilde{y}^*=\lim_{\tilde{\mu}\to\infty}\tilde{\mu}^{-1} = 0$. Use this result with \cref{eq:ngm-f1} to find that $\tilde{x}^*=\lim_{t\to\infty}\tilde{x}(t)$, must satisfies $f(\tilde{x}^*)=\delta\,\tilde{x}^*$.  Rearrange using $f(0) = 0$ to get $\frac{f(\tilde{x}^*) - f(0)}{\tilde{x}^*}=\delta$.  By strict concavity $f'(\tilde{x}^*)<\frac{f(\tilde{x}^*)-f(0)}{\tilde{x}^*}=\delta$.  Finally, \cref{eq:ngm-f2} gives us $G(\tilde{x}^*,\tilde{\mu}(t),\tilde{y}^*)=f'(\tilde{x}^*)-\delta < 0$. The rate of divergence, by \cref{thm:divergence-speed}, is strictly faster than $r$.

The results of \cref{thm:divergence-speed} only partially characterizes the set of non-solutions---concentrating on the primary case which makes shooting methods difficult and our own methods successful.  Other cases which pose no challenge for shooting methods---such as a divergent $\vxh(t)$ with stationary $\vmuh(t)$ tend to be problem dependent and rely on \cref{ass:interior} (see standard saddle-path analysis in \cite{acemoglu2008introduction}). For example, in the growth example a divergent $\tilde{x}(t)$ leads negative $\tilde{y}(t)$ to which contradict \cref{ass:interior}---and can be eliminated by adding in extra bounds to the optimization problem.

The key takeaway from these results is that shooting methods, and similar approaches imposing the steady-state at a terminal condition, are inherently sensitive to the choice of $T$ since all non-solutions explode exponentially leading to instability when evaluating the system pointwise at the terminal condition.

\subsection{Key Insight}
The key insight into our methods is that we can use the rapid divergence separating the unique solution from the non-solutions to our advantage.

While \cref{thm:divergence-speed} shows that the pointwise evaluation of transversality at a large $T$ is sensitive, it hints that more global approaches to contrast solutions from non-solutions might be effective.  In particular, many function norms and semi-norms can provide a numerically stable alternative.  If a non-solution diverges, the norm of the corresponding state or co-state variables will eventually exceed that of the optimal solution for a large enough $T$, thereby violating \cref{eq:f4}. Therefore, any algorithm that solves \cref{eq:f1,eq:f2,eq:f3,eq:f-iv} while controlling a norm of the state and co-state variables over a finite time horizon can get arbitrarily close to the optimal solution.

 While we could use this insight for different types of function approximation and norms (e.g., \cite{spooky} empirically shows similar results arising from inductive bias in deep learning), kernel methods have the advantage that the norms are precisely defined in the Reproducing Kernel Hilbert Space (RKHS) and are numerically stable even for large $T$.

 {\it Sobolev norm solutions.~} Before discussing function approximation in the RKHS space and determining whether it aligns with the solution concept of the DAE, we can consider the natural function space in which solutions to our problem class reside.  This provides us with a precise way to compare degrees of divergence that is not pointwise.

Differential equations of this sort typically align with a space defined by the Sobolev norm of the solution's derivative.
In particular, we consider the Sobolev-$2,2$ space of functions, which is defined as the space of functions $w(\cdot)$
where the function, as well as its first and second (weak) derivatives are square integrable over the domain $[0, T]$:
\begin{align*}
	\sobolev^{2,2}([0,T]) &= \{ w(\cdot) : w(\cdot), \dot{w}(\cdot), \ddot{w}(\cdot) \in L^2([0,T]) \}.
\end{align*}
For any $w(\cdot) \in \sobolev^{2,2}([0,T])$, we have that $\dot w(\cdot)$ lives in the Sobolev-$1,2$ space-%
the space of square-integrable functions with square-integrable first (weak) derivatives.
The Sobolev-$1,2$ norm of $\dot w(\cdot)$, also defined as the Sobolev-$2,2$ semi-norm of $w(\cdot)$,
can be viewed as a measure of complexity of the differential equation solution $w(\cdot)$:
\begin{align*}
	\vert w(\cdot) \vert_{\sobolev^{2,2}([0,T])}
	:= \Vert \dot w(\cdot) \Vert_{\sobolev^{1,2}([0,T])}
	:= \left( \int_0^T \left( \vert \dot{w}(t) \vert^2 + \vert \ddot{w}(t) \vert^2 \right) dt \right)^{1/2}.
\end{align*}
%


In other words: solutions (and non-solution on a finite domain), and their derivatives belong to a Sobolev space with a well-defined semi-norm.%
\footnote{
	We note that the semi-norm $\vert w(\cdot) \vert_{\sobolev^{2,2}}$ is a more natural measure of complexity than the norm $\Vert  w(\cdot) \Vert_{\sobolev^{2,2}}$.
	Consider for example a problem where $w(T)$ converges to some steady state $w(\infty)$ as $T \to \infty$.
	Convergence to a steady state implies that $\dot w(T) \to 0$ as $T \to \infty$, and so it is possible for $\Vert \dot w(\cdot) \Vert_{L^2([0,T])}$ to be bounded.
	However, if $w(\infty) > 0$, then $\Vert  w(\cdot) \Vert_{\sobolev^{1,2}([0,T])}$ will diverge as $T \to \infty$.
	In other words, norms of $w(\cdot)$ are sensitive to the scaling and location of the steady state,
	while norms of $\dot w(\cdot)$ are not.
	}
These are reasonable assumptions given the connections between infinite-horizon optimal solutions in economic growth models and Sobolev spaces; see \citet{van2007optimal} and \citet{CHICHILNISKY1977504}.

Back to our setup: intuitively, solutions to \cref{eq:f1,eq:f2,eq:f3,eq:f-iv} either converge and fulfill \cref{eq:f4} or diverge, for example with $\dot{\tilde{\vmu}}(t) > 0$.  This would lead to solutions having a lower semi-norm. As $T$ increases, the solution and non-solutions increasingly separate since \cref{thm:divergence-speed} shows $\frac{\dot{\tilde{\vmu}}^{(m)}(t)}{\tilde{\vmu}^{(m)}(t)} \;\geq\; r$, for some $m = 1,\cdots,M$. Therefore, given the bounding in \cref{ass:interior} to find the solution that fulfills transversality, it is sufficient to choose among all non-solutions those where $\Vert \vx \Vert_{\sobolev^{1,2}([0,T])} + \Vert \vmu \Vert_{\sobolev^{1,2}([0,T])}$ is minimized.

	\section{Method}
	\label{sec:method}
	In this section we first define our approximation class and present an algorithm for solving an underdetermined DAEs, \cref{eq:f1,eq:f2,eq:f3,eq:f-iv},  using ridgeless kernel regression. Next,  \cref{thm:bound-norms} shows that a minimum-norm solution is a sufficient condition to satisfy the pointwise transversality requirement---ensuring that the unique solution to the underdetermined kernel regression is the one fulfilling transversality.  Finally, we establish in \cref{thm:consistency} that ridgeless kernel regression is consistent and therefore asymptotically enforces the minimum norm condition and, by \cref{thm:bound-norms}, solves the full system \cref{eq:f1,eq:f2,eq:f3,eq:f4,eq:f-iv}.  Although intuitive, the result is nontrivial because it requires demonstrating alignment and mathematical consistency between the RKHS of the approximate solution, its derivatives, and the solution concept of the DAE itself.

\subsection{Kernel Method}

Our algorithm models $\vx(t)$, $\vmu(t)$, and $\vy(t)$, together with their derivatives, using kernel machines---an exemplar-based approximation in which functions are expressed as combinations of their values at the data, weighted by a function that encodes a notion of distance \citep{probml1}.
More specifically, our approximation ensures that the DAE is satisfied on some finite set of (possibly irregular) points $\train := \{t_1,\ldots,t_N\}$,
which we refer to as the ``training data,''
while guaranteeing that it has the minimum function norm amongst all possible solutions satisfying this condition.
As a non-parametric method, the number of parameters defining our approximation grows with the data.
Given a kernel function $k(\cdot,\cdot)$, our approximations take the form:
\begin{equation}
	\begin{gathered}
		\vdotxh(t) = \tsum_{j=1}^N \balpha^x_j k(t, t_j), \qquad
		\vdotmuh(t) = \tsum_{j=1}^N \balpha^\mu_j k(t, t_j), \qquad
		\vdotyh(t) = \tsum_{j=1}^N \balpha^y_j k(t, t_j),\\
		\vxh(t) = \vx_0 + \tint_0^t \dot\vxh(\tau) d\tau, \qquad
		\vmuh(t) = \vmuh_0 + \tint_0^t \dot\vmuh(\tau) d\tau, \qquad
		\vyh(t) = \vyh_0 + \tint_0^t \dot\vyh(\tau) d\tau, \qquad
	\end{gathered}
	\label{eq:kernel}
\end{equation}
where $\balpha^x_j$, $\balpha^\mu_j$, $\balpha^y_j$, $\vmuh_0$, and $\vyh_0$ are parameters fit to fulfill our DAE.\footnote{For a given kernel, the integration in \cref{eq:kernel} can be done in closed form or numerically, i.e. $\int_0^t k(\tau, t_j)\diff\tau$ for each $t_j$.} We  approximate the time-derivative, and integrate to obtain the function values.  While it is possible to approximate the function itself, the benefits of this formulation are that the initial condition $\vxh(0) = \vx_0$ is directly enforced, and extrapolation performance is significantly improved.\footnote{In particular, since most kernels will have $\lim_{t\to\infty}k(t,t_j) = 0$ for any $t_j$, approximating the derivatives leads to $\lim_{t\to\infty}\vmuh(t) = \vmuh_0$ for any fixed $\train$.  Alternatively, if the function itself was approximated then, for any fixed $\train$, $\lim_{t\to\infty}\vmuh(t) =0$.  That said, given that our goal is to solve for only short- and medium-run behavior, and our consistency results in \cref{thm:consistency} only apply on $[0,T]$, approximating the function itself often works in practice.}

In this paper, we use a Matérn kernel for $k(\cdot, \cdot)$, with smoothness $\nu$ and lengthscale $\ell$ (see \cref{def:Matern-kernels} in \cref{sec:norms}). The choice of the Matérn kernel family is driven by theoretical considerations to formally align with the solution concept of the DAE, but many other kernels (e.g., Gaussian kernels) have similar performance in practice, even if they impose more smoothness than is strictly necessary.

{\it Function Norms.~}
Central to these methods is that the kernel function, $k(\cdot,\cdot)$ has an associated function space, its Reproducing Kernel Hilbert Space, $\rkhs$ which provides an inner product and an associated function norm.  Moreover, the norm of functions in this space can be calculated as a quadratic form of its coefficients.  In particular, for the approximated $\vdotxh(t)$ and $\vdotmuh(t)$ in \cref{eq:kernel}, $\Vert \vdotxh^{(m)} \Vert^2_{\mathcal{H}}$ and $\Vert \vdotmuh^{(m)} \Vert^2_{\mathcal{H}}$, are constructed as follows:
\begin{equation}
	\begin{gathered}
		\Vert \vdotxh^{(m)} \Vert^2_{\mathcal{H}} = \sum_{i=1}^N \sum_{j=1}^N \alpha^{x^{(m)}}_i \, \alpha^{x^{(m)}}_j \, k(t_i, t_j), \\
		\Vert \vdotmuh^{(m)} \Vert^2_{\mathcal{H}} = \sum_{i=1}^N \sum_{j=1}^N \alpha^{\mu^{(m)}}_i \, \alpha^{\mu^{(m)}}_j \, k(t_i, t_j)
	\end{gathered}\label{eq:kernel-norms}
\end{equation}
\noindent where superscript $(m)$ denotes the coefficients corresponding to the $m$-th state or co-state variable. For instance, $\balpha^{x^{(m)}}_i$ is the $i$-th element of the learnable coefficients for the $m$-th state variable.

The subscript $\mathcal{H}$ denotes the Reproducing Kernel Hilbert Space (RKHS) associated with the kernel $k(\cdot,\cdot)$. The RKHS norm $\Vert\cdot\Vert_{\mathcal{H}}$ measures the complexity of a function in a way determined by $K$ and is used to control the smoothness of the approximating functions. For more details, see \citet[Chapters~2\ \&\ 3]{smola1998learning}. In the case of Matérn kernels, as will be discussed later, this norm has a concrete smoothness interpretation: it directly controls the magnitude of the function’s derivatives up to a certain order.

{\it Ridgeless Kernel Regression.~}
Kernel methods are tailored to solve empirical risk minimization (ERM) style objective in this function space such as $\min_{h\in\rkhs}\{\sum_{i=1}^N \Phi(h(t_i)) + \lambda ||h||_{\rkhs}^2\}$ for some loss $\Phi(\cdot)$ and regularization term $\lambda > 0$.  The Representer Theorems show this has a unique solution of the form $h(t) = \sum_{j=1}^N \alpha_j k(t, t_j)$ for some $\alpha_j \in \R$ \citep{scholkopf2001generalized,smola1998learning}, with norm $\Vert h \Vert^2_{\rkhs} = \sum_{i=1}^N \sum_{j=1}^N \alpha_i \, \alpha_j \, k(t_i, t_j)$.  Furthermore, standard results (e.g., \citet{Rakhlin-Liang} and \citet{hastie2022surprises}) show that in the limit as $\lambda \to 0$ this is equivalent to directly minimizes that norm subject to interpolating the data, i.e., $\min_{h\in\rkhs} ||h||_{\rkhs}^2 \text{ s.t. }\Phi(h(t_i)) = 0 $ for all $t_i \in \train$.

Mapping to our problem, the loss will interpolate the system of equations in \cref{eq:f1,eq:f2,eq:f3,eq:f-iv} and minimize the sum of RKHS norms of the approximating functions to solve,
\begin{align}
		\min_{\substack{
				\vdotxh, \vdotmuh,\vdotyh
		}}\, & \left(
		\tsum_{m=1}^{\Mx}\, \Vert \vdotxh^{(m)} \Vert^2_{\rkhs} +
		\tsum_{m=1}^{\Mx}\, \Vert \vdotmuh^{(m)} \Vert^2_{\rkhs}\right)\label{eq:erm-norm}\\
		\text{s.t.}~\,
		& \vdotxh(t_i) = \vF(\vxh(t_i), \vmuh(t_i), \vyh(t_i)),\quad\text{for all } t_i \in \train\nonumber\\
		& \vdotmuh(t_i) = r\vmuh(t_i)- \vmuh(t_i)\odot\vG(\vxh(t_i), \vmuh(t_i), \vyh(t_i)),\quad\text{for all } t_i \in \train\nonumber\\
		& \mathbf{0} = \vH(\vxh(t_i), \vmuh(t_i), \vyh(t_i)),\quad\text{for all } t_i \in \train.\nonumber
	\end{align}
The power of kernel methods is that the Representer Theorems shows that this infinite-dimensional non-parametric problem in a function space becomes tractable and finite-dimensional in  $\balpha^x_j$, $\balpha^\mu_j$, $\balpha^y_j$, $\vmuh_0$, and $\vyh_0$ since solutions can be expressed as \cref{eq:kernel,eq:kernel-norms}.\footnote{The primary tradeoffs for kernel methods is that they require pairwise evaluation across all of the data (i.e., $k(t_i, t_j)$ for all $i,j$), and an out-of-data function evaluation require evaluating $k(t,t_i)$ for all data.  Computational methods can solve these challenges approximately up to millions of observables, \cite{NEURIPS2018_27e8e171,NEURIPS2019_01ce8496}, but as with all non-parametric methods it eventually becomes a limitation.  Another consideration is that while in our current applications it is essential, the strong dependence of solutions on the RKHS norms may or may not be a benefit if the solution concept of the underlying problem is poorly aligned with the RKHS induced by the kernel.} When the scale is such that constrained optimizers are insufficient, one can instead solve the optimization problem for a fixed, small regularization penalty using the more standard ERM formulation. For details, see \cref{section-ridge-regression}.

Intuitively, minimizing \cref{eq:erm-norm} finds the function with the minimum RKHS norm among all possible interpolating solutions.  Next we need to show why this minimum norm solution will be the one which fulfill transversality.

\subsection{Transversality}
While the intuition that a minimum function norm solution is sufficient to enforce transversality is intuitive, we need to proceed cautiously to relate the norms of the DAE solution concept to that of the RKHS.  While many different kernels, and associated norms, could be used in practice, we will emphasize Matérn kernels due to their formal connection to Sobolev spaces.

{\it Matérn kernels.~} For functions defined over compact domains,
it is well established that the Mat\'ern RKHS with $\nu = (P - 1/2)$ and a specific value of $\ell$ is
exactly equal to the $\sobolev^{P,2}([0, T])$ norm for all $P \geq 1$
(see \cref{sec:norms}).
Thus, modelling the derivatives with the $\nu = 1/2$ Mat\'ern kernel (with appropriate lengthscale) produces minimum $\sobolev^{1,2}([0, T])$ derivatives (and thus minimum $\sobolev^{2,2}([0, T])$-seminorm solutions).
We are assuming more curvature than is strictly necessary, given that the solution concept of the DAE only requires a first derivative.  However, targeting the $\sobolev^{2,2}([0, T])$-semi-norm provides the necessary control to ensure transversality, as we will now show.


Given that \cref{eq:f4} is expressed in terms of the product $\vx(t)\odot \vmu(t)$, but that \cref{eq:erm-norm} and Representer Theorems are written in terms of function norms, we first need to show how penalized norms are sufficient to control the $\vx(t) \odot \vmu(t)$ globally---and consequently will bound the pointwise transversality condition.

\begin{restatable}{theorem}{thm:bound-norms}{Bounding with Norms}
Let $f$ and $g$ be elements of $\sobolev^{2,2}([0, T])$.  Then, for some $D_1 < \infty$,
\begin{align}
	\sup_{t} \left| f(t) g(t) \right| &\leq D_1 \left(  \left\Vert f\right\Vert_{\rkhs}^2 + \left\Vert g \right\Vert_{\rkhs}^2 \right),\label{eq:bound-1}\\
	\intertext{and from the Sobolev embedding theorem, there exists some $D_2 < \infty$ such that,}
	\sup_{t} \left| f(t) g(t) \right| &\leq D_2 \left(  \left\Vert \dot{f}\right\Vert_{\rkhs}^2 + \left\Vert \dot{g} \right\Vert_{\rkhs}^2 \right),\label{eq:bound-2}
\end{align}
\noindent where $\rkhs$ is the Mat\'ern RKHS with $\nu = \frac{1}{2}$
\end{restatable}
\begin{proof}
See \cref{sec:boundnorms}.
\end{proof}
This theorem connects the norm of the derivatives to the maximum value the product of two functions can obtain.  Intuitively, this result shows why controlling the RKHS norms of the sum of the derivatives might rule out unbounded and explosive functions. Functions that diverges pointwise on $[0,T]$ must have an RKHS norm that becomes unbounded as $T$ grows.

To reiterate, this shows that a minimum norm of the approximated derivatives is sufficient to control the pointwise transversality condition and aligns with the underlying Sobolev norm of the solution concept of the DAE.  While we have assumed more smoothness than is strictly necessary, e.g., $\sobolev^{2,2}([0, T])$ instead of $\sobolev^{1,2}([0, T])$, this is not a significant limitation in practice as we demonstrate in \cref{sec:results}.\footnote{However, this also shows where these methods are likely to break down.  For example, in cases which are not twice-differentiable almost everywhere, or where the level of the function rather than just its derivatives are essential for selecting the trajectory fulfilling \cref{eq:f4}.}

\subsection{Consistency}
While applying \cref{thm:bound-norms} shows that the min norm solution of the DAE will be the one which fulfills \cref{eq:f4}, we need to show that our kernel solution minimizing  \cref{eq:erm-norm} approximates the true solution despite only satisfying the DAE on a finite number of points.
To that end, we demonstrate that our approximation is a \emph{consistent estimator} of the true minimum norm solution,
meaning that as $N \to \infty$ the empirical solutions $\vxh_N, \vmuh_N, \vyh_N$ converge to \cref{eq:f1,eq:f2,eq:f3,eq:f-iv,eq:f4}.
\begin{restatable}{theorem}{thm:consistency}{Consistency}
Given some $0 < K < \infty$, let $\sS$ be the set of functions $(\vx, \vmu, \vy)$
that satisfy \cref{eq:f1,eq:f2,eq:f3,eq:f-iv} and \cref{ass:regularity}
with $\vx(0) = \vx_0$
and $\Vert \vmu(0) \Vert_\infty, \Vert \vy(0) \Vert_\infty \leq K$.
Then the minimum norm solution
$$
(\vx^*, \vmu^*, \vy^*) =
\textstyle{\inf_{(\vx, \vmu, \vy) \in \sS}}
 \tsum_{m=1}^{\Mx}\, \Vert \dot{x}^{(m)} \Vert^2_\rkhs
+ \tsum_{m=1}^{\Mx}\, \Vert \dot{\mu}^{(m)} \Vert^2_\rkhs
+ \tsum_{m=1}^{\My}\, \Vert \dot{y}^{(m)} \Vert^2_\rkhs
$$
exists and has bounded $\rkhs$ norm.
Moreover,
if $t \in \train$ are drawn uniformly i.i.d. from $[0, T]$ then
the solutions $\vxh_N, \vmuh_N, \vyh_N$ from \cref{eq:erm-norm}
with the Mat\'ern-$1/2$ kernel
satisfies \cref{eq:f1,eq:f2,eq:f3,eq:f-iv} almost everywhere in the limit as $N \to \infty$
and
\[
\begin{aligned}
	\textstyle{\lim_{N \to \infty}}
	& \tsum_{m=1}^{\Mx}\, \Vert \hat{\dot{x}}^{(m)}_N \Vert^2_\rkhs
	+ \tsum_{m=1}^{\Mx}\, \Vert \hat{\dot{\mu}}^{(m)}_N \Vert^2_\rkhs
	+ \tsum_{m=1}^{\My}\, \Vert \hat{\dot{y}}^{(m)}_N \Vert^2_\rkhs
	\\
	\overset{\mathrm{a.s.}}{=} \:
	\textstyle{\inf_{(\vx, \vmu, \vy) \in \sS}}
	& \tsum_{m=1}^{\Mx}\, \Vert \dot{x}^{(m)} \Vert^2_\rkhs
	+ \tsum_{m=1}^{\Mx}\, \Vert \dot{\mu}^{(m)} \Vert^2_\rkhs
	+ \tsum_{m=1}^{\My}\, \Vert \dot{y}^{(m)} \Vert^2_\rkhs.
\end{aligned}
\]
\end{restatable}
\begin{proof}
See \cref{sec:consistency}.
\end{proof}
Though this proof borrows techniques from the statistical learning literature on kernel methods \citep[e.g.,][]{wainwright2019high}, it is non-trivial for several reasons.
First, we need to show that solutions to the DAE exist in the RKHS induced by the Mat\'ern kernel,
and that a minimum norm solution exists.
Second, we need to prove that uniform convergence of derivatives implies uniform convergence of the DAE solution functions themselves.
Given these results, we can be confident that with enough data and a large enough $T$, the min semi-norm solution exists, will be consistently approximated by the empirical solutions $\vxh_N, \vmuh_N, \vyh_N$ fulfilling \cref{eq:f1,eq:f2,eq:f3,eq:f-iv}, and that it will be sufficient to ensure the transversality condition \cref{eq:f4} holds.

	\section{Results}
	\label{sec:results}
	We solve two standard baselines in dynamic economics: the neoclassical growth model in \cref{sec:neoclassical-growth-model}, and a model of
risk-neutral asset pricing in \cref{sec:asset-pricing}. These problems are chosen because they are standard examples in
textbooks \citep[e.g.,][]{ljungqvist2018recursive}, they admit reference solutions
from classical methods, and they have
established results regarding the set of solutions to the
ill-posed versions without the asymptotic boundary conditions. In \cref{sec:neoclassical-growth-model-concave-convex}, we present a case with multiple steady states—a challenging setting where our methods are particularly useful—and summarize additional experiments in \cref{sec:other-examples}.

In all cases, we use a Mat\'ern kernel with $\nu = \frac{1}{2}$, $\ell = 10$, and $\sigma = 1$ (see \cref{def:Matern-kernels}) and use $\train := \{0,1,2,\ldots,40\}$.  Where possible we compare the relative errors of a kernel solution to a benchmark obtained with classic methods, e.g., $\varepsilon_{\vw(t)} := \left|\tfrac{\vwh(t) - \vw(t)}{\vw(t)}\right|$ with corresponding definitions for the other variables. In our baseline model, all cases are solved using open-source constrained optimizers and execute in less than a second.

\subsection{Neoclassical Growth Model}\label{sec:neoclassical-growth-model}
In this section, we solve the neoclassical growth model (also known as the Ramsey–Cass–Koopmans model) summarized by \cref{eq:ngm-f1,eq:ngm-f2,eq:ngm-f3,eq:ngm-f4,eq:ngm-iv}.  Our baseline parameters are $x_0 = 1.0, \delta = 0.1, r = 0.11, f(x) = x^a$ and $a= \frac{1}{3}$.

\begin{figure} 
	\centering
	\includegraphics[width=\textwidth]{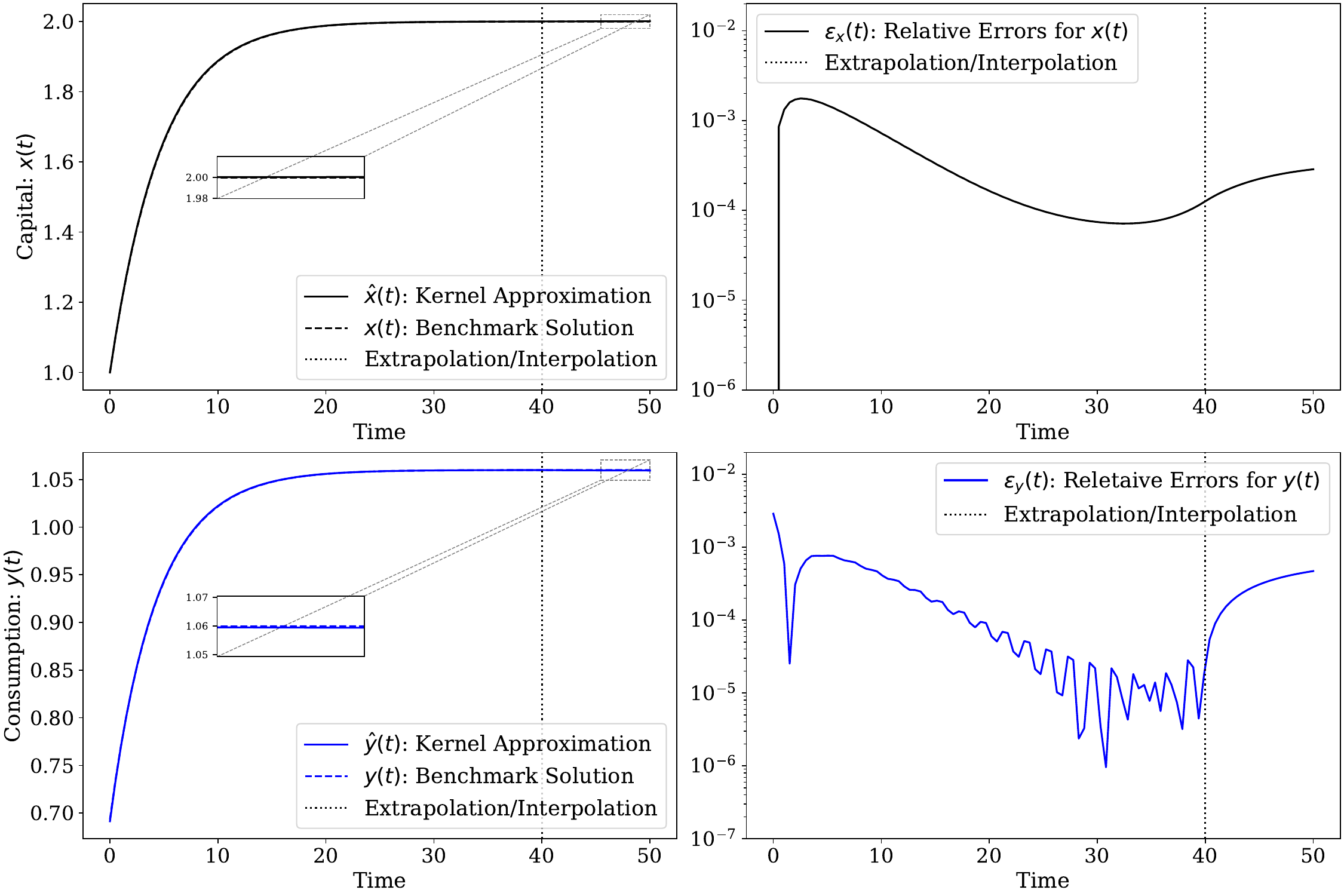}
	\vspace{-7mm}
	\caption{Solution of the neoclassical growth model (\cref{eq:ngm-f1,eq:ngm-f2,eq:ngm-f3,eq:ngm-iv})
		without imposing the transversality condition (\cref{eq:ngm-f4}).  The vertical dotted lines marks the end of the training set.  The top panels show the results for capital, $x(t)$ and associated relative error, and the bottom two show results for consumption, $y(t)$.  The relative error, are low, on the order of $0.01\%$ even when extrapolating beyond $\train$.}
	\label{fig:ngm-results}
\end{figure}

{\it Results.~}
\cref{fig:ngm-results} shows the consumption and capital relative to a benchmark.\footnote{In this case we solved this as a mixed initial-boundary value problem using the analytically calculated boundary condition as a boundary condition. As discussed in \cref{sec:setup}.} The kernel approximation recovers the optimal solution almost perfectly, fulfilling \cref{eq:ngm-f4} despite not being provided the steady-state as a boundary condition.

The vertical dotted lines mark the end of the training set. While not our primary goal, the solution method extrapolates accurately, allowing it to learn the steady state.  As discussed in \cref{sec:method}, this is possible since we approximate the derivatives rather than the functions themselves and Matérn kernels are zero-reverting.

{\it Sufficiency of the minimum norm solution.~}  As discussed in \cref{sec:explosive-solutions}, as long as $f(x)$ is monotone and strictly concave, $f(0) = 0$, and $\delta > 0$, then this model fulfills \cref{thm:divergence-speed}.  That is, non-solutions of $\tilde{\mu}(t)$ diverge asymptotically at a rate greater than $r$.  Since these are the key failures of transversality, the only modification to the optimization problem \cref{eq:erm-norm} required in practice was to impose the bound \( \hat{y}_0 > 0 \), which prevented all other violations of \cref{ass:interior}.\footnote{In general as problems get larger it is helpful to add additional non-negativity or box-bounding constraints from \cref{ass:interior} to the optimization problem minimizing \cref{eq:erm-norm}.  While these will often be non-binding in the optimal solution, they can help optimizers converge.}


{\it Robustness.~}
While \cref{thm:consistency}  demonstrate consistency, it is helpful to check our methods' sensitivity to hyperparameters and features of $\train$:
Section 1.1 of the Supplemental Appendix shows that our methods perform well with a much sparser and irregular $\train$; 
section 1.2 of Supplemental Appendix indicates low sensitivity to different kernel hyperparameters; 
and section 1.3 of Supplemental Appendix demonstrates that the approximation remains effective in the short to medium term, even if $\train$ does not contain large time values, 
which correspond to the solution getting very close to the steady state—an important consideration for problems where the convergence rate is unknown and the appropriate choice of $T$ is not clear a priori.

\subsection{Neoclassical Growth Model with Multiple Steady-States}\label{sec:neoclassical-growth-model-concave-convex}
We now turn to a more complex version of the neoclassical growth model where
$f(x) := A\max\{x^a,b_1 x^a-b_2\}$, as in \cite{azariadis1990threshold,skiba} where $A = 0.5, b_1 = 3.0,$ and $b_2 = 2.5$. The derivative of the production function, $f'(x)$,  exhibits a discontinuity at $\bar{x} = \left(\frac{b_2}{b_1-1}\right)^{\frac{1}{a}}$. As a result this problem has two steady states, but a unique transition path for any given initial condition.

%
This poses a significant challenge for classical algorithms, such as shooting and BVP methods, because the algorithms rely on analytic characterization of the steady-state value to impose the correct boundary-value for a particular initial condition $\vx_0$.  In practice, the set of steady-states and the partitioning of the initial conditions into domains of attraction are not known a-priori and are rarely computed outside of simple problems.

\begin{figure}[H]
	\centering
	\includegraphics[width=\textwidth]{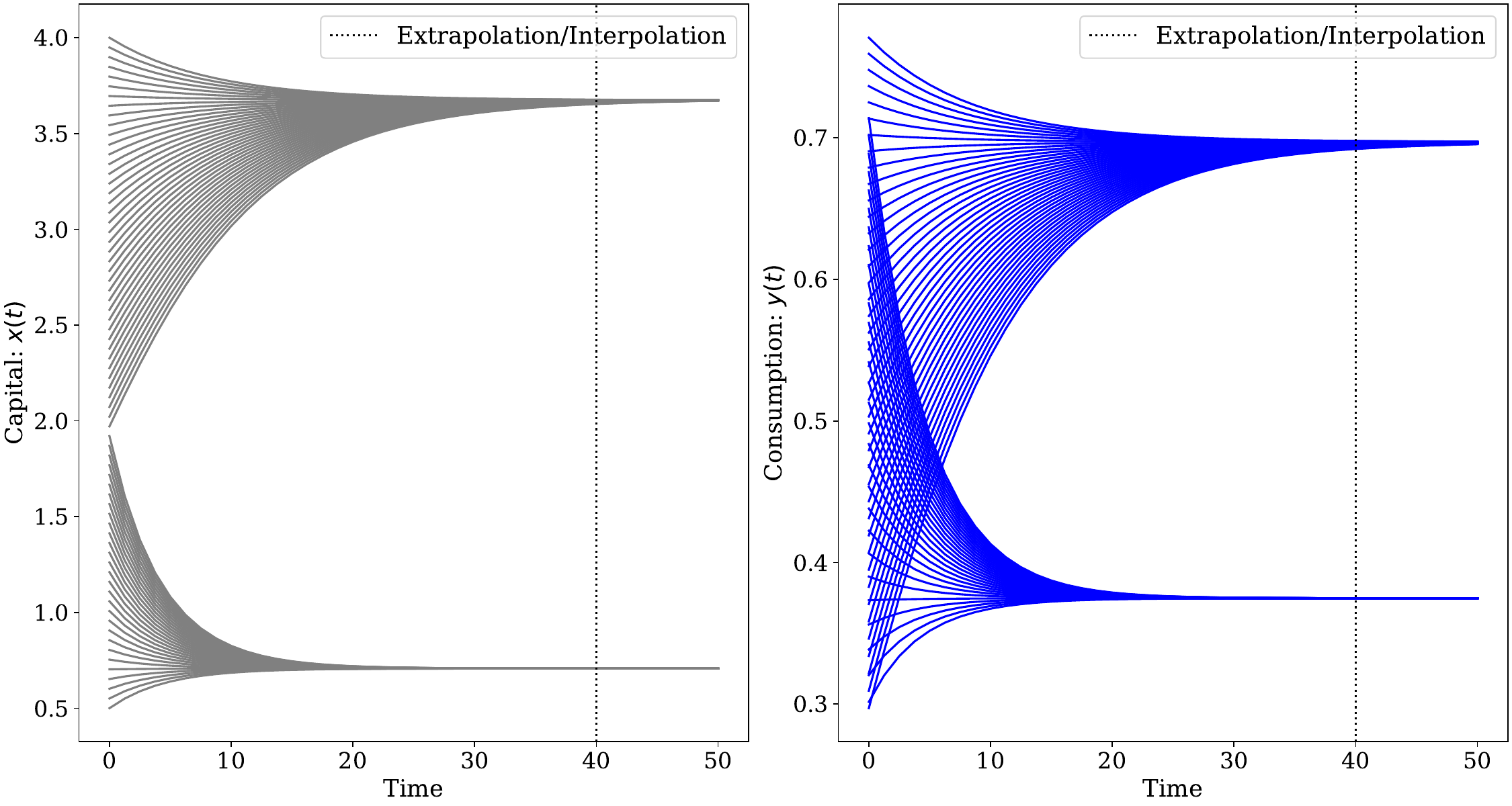} 
	\vspace{-7mm}
	\caption{Solution of the
		neoclassical growth model (\cref{eq:ngm-f1,eq:ngm-f2,eq:ngm-f3,eq:ngm-iv}) with multiple steady states due to the concave-convex production function in \cref{sec:neoclassical-growth-model-concave-convex}.  The left panel shows the solution trajectories for capital, $x(t)$, for $70$ different initial conditions $x_0\in  [0.5,4]$. The right panel shows the solution trajectories for consumption, $y(t)$.
	}
	\label{fig:neoclassical_growth_model_concave_convex_threshold}
\end{figure}

{\it Results.~}
\cref{fig:neoclassical_growth_model_concave_convex_threshold} shows
the results using Mat\'ern kernels for 70 different initial conditions
across the basins of attraction for the two steady states. 

While \cref{thm:divergence-speed} does not hold globally, it is true local to both steady-states---eliminating the key non-solutions.  However, in this case there is an additional concern that an approximate solution might jump into a different basin of attraction and go to the wrong steady-state.  While possible, we found that all of the examples converge to the ``correct'' steady state; i.e. the steady state that correctly corresponds
with the supplied initial condition $x_0$ despite not being provided the set of steady-states or their basins of attraction.

Intuitively, this behavior is also consequence of the minimum norm solution.  Consider the two possible trajectories $x_0$ to each of the two steady states: the trajectory with smaller gradients and less steep dynamics will have a smaller norm.

\subsection{Linear Asset Pricing}\label{sec:asset-pricing}
Models of asset pricing and rational bubbles are relatively simple and often admit closed-form solutions. These models have traditionally served a pedagogical role in exploring transversality conditions; for instance, see \cite{ljungqvist2018recursive}. In this context, the transversality condition is also referred to as the ``no-bubble'' condition.\footnote{For
	asset pricing models, see \cite{blanchard1982bubbles,DibaGrossman},
	which characterizes the set of solutions not fulfilling transversality
	and connects them to economic bubbles.}

{\it Model.~}
The model values a stream of dividends, where a ``bubble'' is defined as a price path whose dynamics cannot be explained by the dividends process.
 
Let
$x(t) \in \R$ be the flow payoffs from a claim to an asset, and
$\mu(t) \in \R$ be the price of a claim to that asset. For simplicity,
we assume that the flow payoffs, $x(t)$, follow a deterministic linear process.
For a given $x_0$, the key equations are
\begin{align}
	\dot{x}(t) &= c + g x(t), \label{eq:asset-pricing-x-dot}\\
	\dot{\mu}(t) &= r \mu(t) - x(t) := r \mu(t) - \mu(t) \frac{x(t)}{\mu(t)},\label{eq:asset-pricing-mu-dot}\\
	0 &= \lim_{t\rightarrow \infty} e^{-r t}\mu(t)x(t),\label{eq:asset-pricing-no-bubble}
\end{align}
where $c$ and $g$ are constants governing the dividend process, and $r>0$ denotes the discount rate of a risk-neutral investor. \cref{eq:asset-pricing-no-bubble} is the ``no-bubble" condition.
The set of solutions to \cref{eq:asset-pricing-mu-dot,eq:asset-pricing-x-dot},
without imposing \cref{eq:asset-pricing-no-bubble}, can be found analytically:
\begin{align}
	\mu(t) &= \tint_0^{\infty} e^{-r \tau} x(t+\tau) \diff \tau
	= \mu_f(t) + \zeta e^{r t},\label{eq:asset-pricing-ode-general}\\
	\mu_f(t) &:= \tfrac{c}{r-g} + \left(x_0 - \tfrac{c}{r-g}\right)e^{(r-g)t} ,\label{eq:asset-pricing-fundamental}
\end{align}
where $\mu_f(t)$ is interpreted as the ``fundamental'' price of the asset,
and $\zeta \geq 0$ is indeterminate. However, when the ``no-bubble" condition
(i.e., \cref{eq:asset-pricing-no-bubble}) is imposed, this problem is well-posed,
with a unique solution of $\mu(t) = \mu_f(t)$ (i.e., $\zeta = 0$).

{\it Selecting the ``no-bubble'' solution.}
A key advantage of this example is that
\cref{eq:asset-pricing-ode-general}
characterizes the full set of deterministic non-solutions to \cref{eq:asset-pricing-x-dot,eq:asset-pricing-mu-dot}
which do not impose \cref{eq:asset-pricing-no-bubble}. For a given function
norm, apply the triangle inequality to the set of solutions from
\cref{eq:asset-pricing-ode-general} to yield
$\Vert\mu_f\Vert_{\sobolev} \leq \Vert\mu\Vert_{\sobolev} \leq \Vert\mu_f\Vert_{\sobolev} + \zeta \Vert e^{rt}\Vert_{\sobolev}$.  Differentiating yields $\Vert\dot{\mu_f}\Vert_{\sobolev} \leq \Vert\dot{\mu}\Vert_{\sobolev} \leq \Vert\dot\mu_f\Vert_{\sobolev} + \zeta r \Vert e^{rt}\Vert_{\sobolev}$ where $\sobolev$ is a norm such as $\sobolev^{1,2}([0, T])$.  Finally, note that the this semi-norm is minimized when $\zeta = 0$.


To verify the divergence rate in \cref{thm:divergence-speed}, note from \cref{eq:asset-pricing-mu-dot} that if $\vx(t)$ converges to a finite steady state while $\vmu(t)$ diverges, then $\lim_{t\to\infty}\vG(\vx(t),\vmu(t))=\lim_{t\to\infty}\frac{\vx(t)}{\vmu(t)}=0$, so $\dot{\vmu}(t)/\vmu(t)\to r$ and $\vmu(t)$ grows asymptotically at rate $r$. However, note that the speed of separating solutions from non-solutions is slower than our previous example, where the non-solutions asymptotically diverged at rates strictly faster than $r$. 



\begin{figure}
	\centering
	\includegraphics[width=\textwidth]{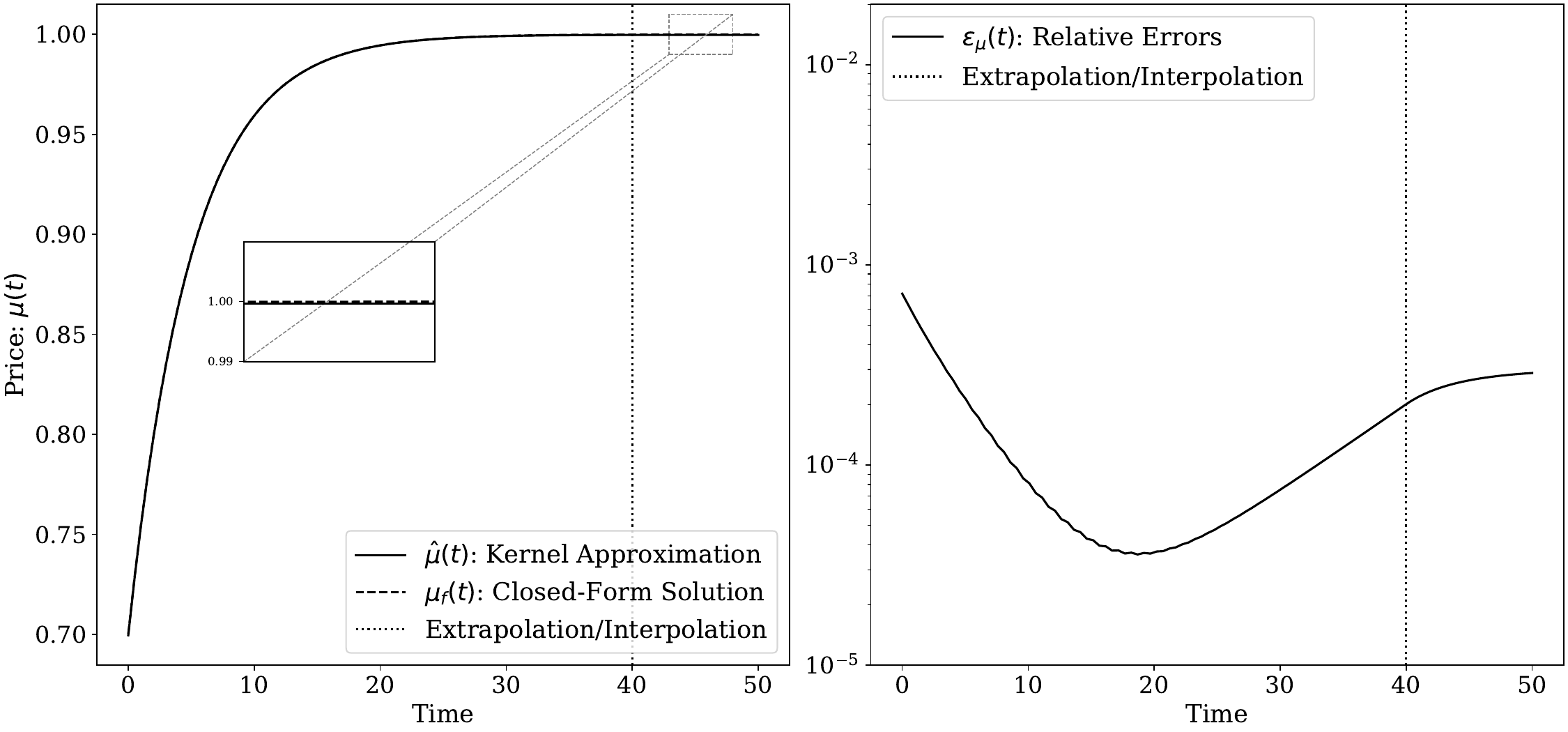}
	\vspace{-7mm}
	\caption{Solution of the linear asset pricing model (\cref{eq:asset-pricing-mu-dot,eq:asset-pricing-x-dot}) without imposing the “no-bubble” condition (\cref{eq:asset-pricing-no-bubble}). The vertical dotted lines marks the end of the training set.}
	\label{fig:asset_pricing_contiguous}
\end{figure}

{\it Results.~}
Our baseline parameters are $x_0 = 1.0, c = 0.02, g = -0.2$, and $r = 0.1$.  \cref{fig:asset_pricing_contiguous} shows the results of the ridgeless kernel machine alongside the fundamental price from \cref{eq:asset-pricing-fundamental}.  Even without imposing the long-run “no-bubble” condition in \cref{eq:asset-pricing-no-bubble}, the kernel method recovers the ``fundamental" price with high accuracy.  As before, the minimum norm  still selects the correct solution and as $T$ grows the norms of non-solutions will grow quickly due to \cref{thm:divergence-speed}.

\subsection{Other Examples}\label{sec:other-examples}
To show that our method can handle problems of increasing dimensionality, we test it on a standard model of human capital and economic growth in section 2.1 of the Supplemental Appendix. In our formulation of the model, there are two jump variables and three co-state variables. With classical methods such as shooting, finding the optimal solution requires a five-dimensional search. By comparison, our algorithm computes the solution in under a second.
To demonstrate the applicability of our algorithm beyond macro and finance, in section 2.2 of the Supplemental Appendix, we test it on the optimal advertising model, a classic framework in the marketing literature.

\subsection{Future Directions}\label{sec:Future}

There are several natural directions for future applications and extensions. One is extending kernel methods to handle inequality and complementarity constraints, which would make them applicable to models such as lifecycle consumption and macroeconomic models with financial frictions. Another direction is adapting kernel methods to stochastic and recursive settings, where conditions like transversality must hold globally and involve expectations over random trajectories. Finally, it is important to study the performance of kernel methods in very high-dimensional settings, such as those found in the trade and spatial economics literature. 

The complexity of kernel methods depends on sample size rather than the dimensionality of the state space. If a kernel captures similarity in the state space, only a modest number of samples may suffice to solve dynamic economic problems, making these methods well-suited for both very high-dimensional and recursive stochastic models.




	\section{Conclusion}
	\label{sec:Conclusion}
	This paper introduces ridgeless kernel methods for solving deterministic, infinite-horizon, continuous-time dynamic models formulated as DAEs. For a broad class of models, we show that selecting the minimum-norm solution guarantees transversality, addressing the main computational obstacle of traditional approaches. Kernel methods are natural for this task, since their function norms are explicit, allowing us to establish sufficiency results linking kernel solutions to the DAE solution concept.

	\appendix
	\renewcommand{\thesection}{Appendix \Alph{section}}
	
	\section{Ridge Regression Formulation}
\label{section-ridge-regression}
The optimization problem in \cref{eq:erm-norm} is equivalent to the limit of ridge regression, as follows:  
\begin{align*}
	&\lim_{\lambda \to 0} \Bigg\{ \min_{\substack{\vdotxh, \, \vdotmuh, \, \vdotyh}} \; \Bigg\{
	\sum_{t_i \in \train} \Bigg[
	\left\Vert \vdotxh(t_i) - \vF(\vxh(t_i), \vmuh(t_i), \vyh(t_i)) \right\Vert_2^2 \\
	+ &\left\Vert \vdotmuh(t_i) - r \, \vmuh(t_i) + \vmuh(t_i) \odot \vG(\vxh(t_i), \vmuh(t_i), \vyh(t_i)) \right\Vert_2^2
	+ \left\Vert \vH(\vxh(t_i), \vmuh(t_i), \vyh(t_i)) \right\Vert_2^2 \Bigg] \\
	&\quad \quad + \left\Vert \vxh(0) - \vx_0 \right\Vert_2^2
	+ \lambda \Big(
	\sum_{m=1}^{\Mx} \Vert \vdotxh^{(m)} \Vert^2_{\rkhs}
	+ \sum_{m=1}^{\Mx} \Vert \vdotmuh^{(m)} \Vert^2_{\rkhs}
	\Big) \Bigg\} \Bigg\},
\end{align*}  

The formulation in \cref{eq:erm-norm} is referred to as \emph{ridgeless} due to the vanishing ridge penalty term as \( \lambda \to 0 \). In practice, rather than taking the limit, one can simply choose a small fixed \( \lambda \). For the applications presented in this paper, we solved the problem for \( \lambda \) in the range \(10^{-4}\) to \(10^{-6}\). The results are omitted, as they were nearly identical to the solution of the optimization problem in \cref{eq:erm-norm}.

	\section{Connection Between Sobolev-${P,2}$ Spaces and Mat\'ern RKHS}
\label{sec:norms}

\begin{definition}[Matern Kernel]
	\label{def:Matern-kernels}
Let $k_{\nu,\ell}(\cdot, \cdot): \R \times \R \to \R$ denote the Mat\'ern covariance function
with smoothness $\nu$ and lengthscale $\ell$.
For the purposes of this paper, we will define $k_{\nu,\ell}$ as:
\begin{equation}
	k_{\nu,\ell}(t, t') = \kappa(\vert t - t' \vert),
	\qquad \widehat{\kappa}(\omega) := \left( 1 + \ell^2\omega^2 \right)^{-\nu - 1/2},
	\label{eq:matern}
\end{equation}
where $\widehat{(\cdot)}$ corresponds to the Fourier transform.
\end{definition}
Note that \cref{eq:matern} corresponds to the standard Mat\'ern kernel definition \citep{rasmussen2006gaussian} after appropriately scaling the inputs and outputs.
$\nabla^{(r)}k_{\nu,\ell}(\cdot, \cdot)$ will denote the $r^\mathrm{th}$ derivative of $k_{\nu,\ell}$ with respect to its its first argument.

Given some interval $[0, T] \subseteq \R$,
$\rkhs^{\nu,\ell}([0, T])$ denotes the RKHS of $[0, T] \to \R$ functions where the reproducing kernel
is equal to $k_{\nu,\ell}$.
$\sobolev^{P,2}([0, T])$ denotes to the Sobolev-$P,2$ space
of $[0, T] \to \R$ functions.
Whenever possible we will drop the superscripts for $\rkhs^{\nu,\ell}([0, T])$.

\begin{theorem}[Equivalence of $\sobolev^{P,2}$ and $\rkhs^{\nu,\ell}$]\label{propo:A1}
	For any positive integer $P$, any $\ell > 0$, and any $\nu = P - 1/2$
	there exists positive constants $C_1(\nu, \ell)$ and $C_2(\nu, \ell)$ so that
	\[
	C_1(\nu, \ell) \Vert w \Vert_{\sobolev^{P,2}([0, T])}
	\leq \Vert w \Vert_{\rkhs^{\nu,\ell}([0, T])}
	\leq C_2(\nu, \ell) \Vert w \Vert^{\sobolev^{P,2}([0, T])}.
	\]
	for all $w \in \sobolev^{P,2}$.
	In other words, the Sobolev norm $\Vert \cdot \Vert_{\sobolev^{P,2}([0,T])}$
	and the RKHS norm $\Vert \cdot \Vert_{\rkhs^{\nu,\ell}([0, T])}$ are equivalent,
	and thus $\sobolev^{P,2}([0, T]) = \rkhs^{\nu,\ell}([0, T])$.
	\label{prop:sobolev_matern}
\end{theorem}
\begin{proof}
	We begin by first establishing an equivalence between
	$\sobolev^{P,2}(\R)$ and $\rkhs^{\ell,\nu}(\R)$.
	In the Fourier domain, the Sobolev norm for $\R \to \R$ functions is given by
	\[
	\Vert w \Vert_{\sobolev^{P,2}(\R)} = \left\Vert \widehat w(\cdot) \left( 1 + ( \cdot )^2 \right)^{P/2} \right\Vert_{L_2(\R)},
	\]
	and, for any RKHS $\rkhs(\R)$ with stationary reproducing kernels, the RKHS norm for $\R \to \R$ functions is given by
	\[
	\Vert w \Vert_{\rkhs(\R)} = \left\Vert \widehat w(\cdot) \left( \widehat{\kappa}(\cdot) \right)^{-1/2} \right\Vert_{L_2(\R)}
	\]
	The Mat\'ern kernel, as defined in \cref{eq:matern},
	has a Fourier transform that decays at at a rate of $(1 + \vert\cdot\vert^2)^{-P}$
	and thus $\Vert \cdot \Vert_{\rkhs^{\nu,\ell}(\R)}$
	is bounded above and below by a constant multiple of $\Vert \cdot \Vert_{\sobolev^{P,2}(\R)}$
	where the constant only depends on $\ell$.
	This argument can be generalized to
	prove an equivalence between
	$\sobolev^{P,2}([0, T])$ and $\rkhs^{\ell,\nu}([0, T])$,
	as the domain $[0, T]$ trivially has a Lipschitz boundary.
	See \citep[][Corollary~10.48]{wendland2004scattered} for details.
\end{proof}

\begin{corollary}[Equality of $\sobolev^{P,2}$ and $\rkhs^{\nu,\ell}$ with specific $\nu,\ell$ values]
	For any $P$, there exists a value of $\ell$ so that,
	for all $w \in \sobolev^{P,2}([0, T])$:
	\[
	\Vert w \Vert_{\sobolev^{P,2}([0, T])} = \Vert w \Vert_{\rkhs^{P - 1/2,\ell}([0, T])}.
	\]
	\label{prop:sobolev_matern_exact}
\end{corollary}
	\section{Proof of ~\cref{thm:bound-norms}}
\label{sec:boundnorms}
\restate{thm:bound-norms}
\begin{proof}
	Let $k(\cdot,\cdot)$ be the kernel associated with the RKHS $\rkhs$ where $\sup_{t}k(t,t) \leq K < \infty$.  For $f,g \in \rkhs$ by the reproducing property,
	\begin{equation}
		f(t) = \langle f,k(t,\cdot) \rangle_{\rkhs} \quad \text{and} \quad g(t) = \langle g,k(t,\cdot)\rangle _{\rkhs}\label{eq:f-g-rep}
	\end{equation}
	Then substitute with \cref{eq:f-g-rep}, use the Cauchy-Schwarz inequality, and then the kernel bound to get
	\begin{align*}
		|f(t) g(t)| &\leq |f(t)| |g(t)|\leq ||f||_{\rkhs} ||k(t,\cdot)||_{\rkhs^{\Mx}}\, ||g||_{\rkhs} ||k(t,\cdot)||_{\rkhs}\\
		&\leq K ||f||_{\rkhs} ||g||_{\rkhs}
		\intertext{Then, by the Arithmetic-Geometric Mean Inequality,}
		&\leq \frac{K}{2}\left( ||f||_{\rkhs}^2 + ||g||_{\rkhs}^2 \right)
		\intertext{By \cref{propo:A1}, there exists a constant $K'$ such that}
		&\frac{K}{2}\left( ||f||_{\rkhs}^2 + ||g||_{\rkhs}^2 \right)\leq K'\left( ||f||_{\sobolev^{1,2}([0, T])}^2 + ||g||_{\sobolev^{1,2}([0, T])}^2 \right)\\
		\intertext{From the Sobolev embedding theorem}
		&  K'\left( ||f||_{\sobolev^{1,2}([0, T])}^2 + ||g||_{\sobolev^{1,2}([0, T])}^2 \right) \leq  K'\left( ||\dot{f}||_{\sobolev^{1,2}([0, T])}^2 + ||\dot{g}||_{\sobolev^{1,2}([0, T])}^2 \right)\\
		\intertext{Finally, from \cref{propo:A1}, there exists some $D$ such that}
		&\leq D \left(  \left\Vert \dot{f}\right\Vert_{\rkhs}^2 + \left\Vert \dot{g} \right\Vert_{\rkhs}^2 \right).
	\end{align*}
\end{proof}
	\section{Proof of Theorem~\ref{thm:consistency}}
\label{sec:consistency}

\restate{thm:consistency}

Throughout this proof we will drop the domain $[0, T]$ from the $\sobolev$ for brevity.
For notational simplicity we redefine $\dot{\vmu}(t) = r \vmu(t) - \vmu(t) \odot \vG(\vx(t),\vmu(t), \vy(t))\equiv \vGh(\vx(t),\vmu(t), \vy(t))$.
It is trivial to see that $\vGh$ is Lipschitz with a Lipschitz first derivative if $\vG$ is.

We break up this proof into a series of lemmas.

\begin{lemma}
	\label{lemma:sobolev_solutions}
	For every $(\vx, \vmu, \vy) \in \sS$, the functions
	$\evx^{(1)}, \ldots, \evx^{(\Mx)}$,
	$\mu^{(1)}, \ldots, \mu^{(\Mx)}$, and
	$\evy^{(1)}, \ldots, \evy^{(\My)}$,
	are all elements of $\sobolev^{2,2}$.
\end{lemma}
\begin{proof}
	First, we note that \cref{ass:regularity} allows us to rewrite the DAE as an ODE.
	Specifically, \cref{eq:f-iv} can be rewritten an equation of $\dot\vy(t)$ by differentiating both sides with respect to $t$
	\begin{align*}
		\vzero &= \nabla_{\vx}\vH(\vx,\vmu,\vy) \dot\vx(t) + \nabla_{\vmu}\vH(\vx,\vmu,\vy) \dot\vmu(t) + \nabla_{\vy}\vH(\vx,\vmu,\vy) \dot\vy(t) \\
		&= \nabla_{\vx}\vH(\vx,\vmu,\vy) \vF(\vx(t), \vmu(t), \vy(t)) + \nabla_{\vmu}\vH(\vx,\vmu,\vy) \vGh(\vx(t), \vmu(t), \vy(t)) \\
		&\:\:\: + \nabla_{\vy}\vH(\vx,\vmu,\vy) \dot\vy(t).
	\end{align*}
	By assumption, $\nabla_{\vy}\vH(\vx,\vmu,\vy)$ is non-singular on all relevant trajectories, so we have that
	\begin{align*}
		\dot\vy(t) &= \nabla_{\vy}\vH(\vx,\vmu,\vy)^{-1} \Bigl( \nabla_{\vx}\vH(\vx,\vmu,\vy) \vF(\vx(t), \vmu(t), \vy(t))
		\\ &\qquad\qquad\qquad\qquad+ \nabla_{\vmu}\vH(\vx,\vmu,\vy) \vGh(\vx(t), \vmu(t), \vy(t)) \Bigr)
		\\ & =: \vC(\vx,\vmu,\vy).
	\end{align*}
	By Lipschitz continuity of $\vF$, $\vGh$, the derivative of $\vH$, the inverse of $\nabla_{\vy}\vH$, and all their respective derivatives,
	we have that $\vC$ is Lipschitz continuous with a Lipschitz first derivative over the interval $[0, T]$.

	We thus can rewrite the DAE as the following ODE system:
	$$
	\dot\vx(t) = \vF(\vx(t), \vmu(t), \vy(t)), \quad
	\dot\vmu(t) = \vGh(\vx(t), \vmu(t), \vy(t)), \quad
	\dot\vy(t) = \vC(\vx(t), \vmu(t), \vy(t)),
	$$
	where $\vF, \vGh, \vC$ are all Lipschitz continuous with Lipschitz first derivatives.
	By Lipschitz continuity of $\vF$, $\vGh$, $\vC$ and the Cauchy-Lipschitz-Picard theorem
	\citep[e.g.][Thm.~7.3]{brezis2011functional},
	we have that every $(\vx, \vmu, \vy) \in \sS$ are continuous and continuously differentiable
	over $[0, \infty)$.
	This fact implies that $\vx, \vmu, \vy$ and $\dot\vx, \dot\vmu, \dot\vy$ are bounded over $[0, T]$.
	Now consider
	$\ddot\vx$,
	$\ddot\vmu$,
	$\ddot\vy$
	\[
	\begin{aligned}
		\ddot{\vx}(t) &= \dot\vF(\vx(t), \vmu(t), \vy(t)) \\
		\ddot{\vmu}(t) &= \dot\vGh(\vx(t), \vmu(t), \vy(t)) \\
		\ddot{\vy}(t) &= \dot\vC(\vx(t), \vmu(t), \vy(t)).
	\end{aligned}
	\]
	Because $\vF$, $\vGh$, and $\vC$ admit Lipschitz first derivatives,
	we have that $\ddot \vx$, $\ddot \vmu$, and $\ddot \vy$ are also bounded over $[0, T]$.
	Since bounded functions are trivially square integrable,
	we have that
	\begin{itemize}
		\item $\Vert \evx^{(1)} \Vert_{\sobolev^{2,2}}, \ldots, \Vert \evx^{(\Mx)} \Vert_{\sobolev^{2,2}} < \infty$,
		\item $\Vert \mu^{(1)} \Vert_{\sobolev^{2,2}}, \ldots, \Vert \mu^{(\Mx)} \Vert_{\sobolev^{2,2}} < \infty$, and
		\item $\Vert \evy^{(1)} \Vert_{\sobolev^{2,2}}, \ldots, \Vert \evy^{(\My)} \Vert_{\sobolev^{2,2}} < \infty$.
	\end{itemize}
\end{proof}

\begin{lemma}
	\label{lemma:set_s_closed}
	Let $M'$ be the total dimension of the DAE (i.e., $M' = 2 \Mx + \My$), and define $\sobolev^{2,2}_M$ as the
	Hilbert space of functions
	$(\vx(\cdot), \vmu(\cdot), \vy(\cdot)): [0, T] \to \R^{M'}$ equipped with the norm:
	\[
	\Vert (\vx(\cdot), \vmu(\cdot), \vy(\cdot)) \Vert_{\sobolev^{2,2}_{M'}}^2
	= \tsum_{m=1}^{\Mx}\, \Vert \evx^{(m)} \Vert^2_{\sobolev^{2,2}}	+
	\tsum_{m=1}^{\Mx}\, \Vert \evmu^{(m)} \Vert^2_{\sobolev^{2,2}} +
	\tsum_{m=1}^{\My}\, \Vert \evy^{(m)} \Vert^2_{\sobolev^{2,2}}.
	\]
	Then the set $\sS$ is closed in $\sobolev^{2,2}_{M'}$.
\end{lemma}
\begin{proof}
	Consider a Cauchy sequence $(\vx_n, \vmu_n, \vy_n) \in \sS$.
	By \cref{lemma:sobolev_solutions},
	we have that $\evx_n^{(m)} \in \sobolev^{2,2}$ for all $m \in [1, \Mx]$.
	By completeness of $\sobolev^{2,2}$,
	we have that $\evx_n^{(m)} \to \evx^{(m)}$ for some $\evx^{(m)} \in \sobolev^{2,2}$;
	i.e. for every $\epsilon$, there exists some $n'$ such that
	$\Vert \evy_{n'}^{(m)} - \evy^{(m)} \Vert_{\sobolev^{2,2}} < \epsilon$.
	\begin{align*}
		\sup_{t \in [0, T]} \Vert \evx^{(m)}_{n'}(t) - \evx^{(m)}(t) \Vert_\infty
		&= \sup_{t \in [0, T]} \left\langle k(t, \cdot), \evx^{(m)}_{n'}(t) - \evx^{(m)}(t) \right\rangle_{\sobolev^{2,2}}
		\\
		&\leq \sup_{t \in [0, T]} \Vert k(t, \cdot) \Vert_{\sobolev^{2,2}} \: \Vert \evx^{(m)}_{n'}(t) - \evx^{(m)}(t) \Vert_{\sobolev^{2,2}}
		\\
		&\leq C \Vert \evx^{(m)}_{n'}(t) - \evx^{(m)}(t) \Vert_{\sobolev^{2,2}} < C \epsilon,
	\end{align*}
	where $k(\cdot, \cdot)$ is the reproducing kernel associated with $\sobolev^{2,2}$ and $C$ is some universal constant.
	The penultimate inequality comes from fact that $\sobolev^{2,2}$ is equivalent to a Mat\'ern RKHS which has a bounded-everywhere reproducing kernel.
	Thus, $\evx^{(m)}_{n'}$ converges uniformly to $\evx^{(m)}$, and so
	\begin{align*}
		\dot\vx(t) = \lim_{n\to\infty} \dot\vx_n(t)
		&= \lim_{n\to\infty} \vF(\vx_n(t), \vmu_n(t), \vy_n(t))
		\\
		&= \mF(\lim_{n\to\infty} \vx_n(t), \lim_{n\to\infty} \vmu_n(t), \lim_{n\to\infty} \vy_n(t))
		= \mF(\vx(t), \vmu(t), \vy(t))
	\end{align*}
	where the penultimate equality comes from the continuity of $\mF$.
	Analogous results hold for $\vmu_n$ and $\vy_n$.
	Moreover, by uniform convergence we have that
	$\Vert \vmu(0) - \vmu_{n}(0) \Vert_\infty$ and
	$\Vert \vy(0) - \vy_{n}(0) \Vert_\infty$
	are arbitrarily small and thus
	$\Vert \vmu(0) \Vert_\infty, \Vert \vy(0) \Vert_\infty < K$.
	Therefore $(\vx, \vmu, \vy) \in \sS$.
\end{proof}

\begin{lemma}
	\label{lemma:min_norm_existance}
	Denote $M'$ and $\sobolev^{2,2}_{M'}$ as in \cref{lemma:set_s_closed}.
	Let $\rkhs_{M'}$ be the Hilbert space of functions $(\dot\vx(\cdot), \dot\vmu(\cdot), \dot\vy(\cdot)): [0, T] \to \R^{M'}$
	equipped with the norm
	\[
	\Vert (\dot\vx(\cdot), \dot\vmu(\cdot), \dot\vy(\cdot)) \Vert_{\rkhs_{M'}}^2
	= \tsum_{m=1}^{\Mx}\, \Vert \dot\evx^{(m)} \Vert^2_{\rkhs} +
	\tsum_{m=1}^{\Mx}\, \Vert \dot\evmu^{(m)} \Vert^2_{\rkhs} +
	\tsum_{m=1}^{\My}\, \Vert \dot\evy^{(m)} \Vert^2_{\rkhs},
	\]
	where $\rkhs$ is the RKHS associated with the Mat\'ern-$1/2$ kernel with some lengthscale $0 < \ell < \infty$.
	Let
	\[
	\begin{aligned}
		B :&=
		{\inf_{(\vx, \vmu, \vy) \in \sS}}
		\Vert (\dot\vx, \dot\vmu, \dot\vy) \Vert_{\rkhs_{M'}}
	\end{aligned}
	\]
	Then there exists some $(\vx^*, \vmu^*, \vy^*) \in \sS$ that achieves this infimum.
\end{lemma}
\begin{proof}
	By \cref{prop:sobolev_matern}, $\sobolev^{2,2}$ is equivalent to the Mat\'ern-$1/2$ RKHS $\rkhs$.
	Note that the elements of $\dot \vw(\cdot)$ for any $\vw(\cdot) \in \sobolev^{2,2}_{M'}$
	are $\sobolev^{1,2}$ functions, and thus
	$$\vw(\cdot) \in \sobolev^{2,2}_{M'} \:\: \Longrightarrow \:\: \dot\vw(\cdot) \in \rkhs_{M'}.$$

	Define the operator $D: \sobolev^{2,2}_{M'} \to \rkhs_{M'}$
	as $D(\vx, \vmu, \vy) = (\dot\vx, \dot\vmu, \dot\vy)$,
	where here $\dot\vx, \dot\vmu, \dot\vy$ denote weak derivatives.
	Note that $D$ is a surjective and bounded linear operator between two Hilbert spaces,
	and that the nullspace of $D$ (i.e. the set of constant functions) is a closed set.
	Since $\sS \subset \sobolev^{2,2}_{M'}$ is a closed subset of a Hilbert space (\cref{lemma:set_s_closed}),
	$\dot\sS := D(\sS) \subset \sobolev^{2,2}_{M'}$ is also closed subset of a Hilbert space \citep[see e.g.][Exercise~2.10]{brezis2011functional}.
	By the existence portion of the Hilbert projection theorem, there exists a (potentially non-unique) minimum $\rkhs_{M'}$-norm element of $\dot\sS$
	(i.e. there exists some $(\dot\vx^*, \dot\vmu^*, \dot\vy^*) \in \dot\sS$ such that
	\(
	\Vert (\dot\vx^*, \dot\vmu^*, \dot\vy^*) \Vert_{\rkhs_{M'}}
	= \inf_{(\vx, \vmu, \vy) \in \dot\sS} \Vert (\dot\vx, \dot\vmu, \dot\vy) \Vert_{\rkhs_{M'}}.
	\)
	We conclude the proof by setting $(\vx^*, \vmu^*, \vy^*)$ to be some element in $\sS$
	such that $(\dot\vx^{*}, \dot\vmu^{*}, \dot\vy^{*}) = D(\vx^*, \vmu^*, \vy^*)$.
\end{proof}

\begin{lemma}
	\label{lemma:rkhs_rademacher}
	For any $0 < C < \infty$,
	define the sets
	\[
	\begin{gathered}
		\sF^C := \{ w \in \rkhs \: : \: \Vert w \Vert_{\rkhs} \leq C \}
		\\
		\tint\sF^C := \{ \tint_{0}^{(\cdot)} w(\tau) d\tau \: : \: w \in \sF \}.
	\end{gathered}
	\]
	Denoting $\widehat\rademacher_N$ as the empirical Rademacher complexity for some dataset $t_1, \ldots, t_N \in [0, T]$,
	we have that
	\[
	\widehat\rademacher_N(\sF^C) \lesssim C N^{-1/2}, \qquad
	\widehat\rademacher_N(\tint\sF^C) \lesssim T C N^{-1/2}.
	\]
\end{lemma}
\begin{proof}
	The Rademacher complexity $\widehat\rademacher_N(\sF^C) \lesssim C N^{-1/2}$ follows a standard result for reproducing kernel Hilbert spaces,
	using the fact that $\rkhs$ (the Mat\'ern-$1/2$ RKHS) has a bounded-everywhere reproducing kernel.
	Bounding the Rademacher complexity of $\tint\sF^C$ mirrors the standard proof of the $\widehat\rademacher_N(\sF^C)$ bound:
	\begin{align*}
		\widehat\rademacher(\tint\sF^C)
		&:=
		\E_{\epsilon_i} \left[ \sup_{w \in \rkhs} \frac 1 N \sum_{i=1}^N \epsilon_i \int_{0}^{t_i} w(\tau) d\tau \right]
		\tag{$\epsilon_i \simiid \mathrm{Rad}$}
		\\
		&=
		\E_{\epsilon_i} \left[ \sup_{w \in \rkhs} \left\langle \frac 1 N \sum_{i=1}^N \epsilon_i \int_{0}^{t_i} k(\tau, \cdot) d\tau,
		\: w(\cdot) \right\rangle_{\rkhs} \right]
		\\
		&\leq
		\E_{\epsilon_i} \left[ \left\Vert \frac C N \sum_{i=1}^N \epsilon_i \int_{0}^{t_i} k(\tau, \cdot) d\tau \right\Vert_{\rkhs} \right]
		\tag{Cauchy-Schwarz inequality}
		\\
		&\leq
		\sqrt{ \E_{\epsilon_i} \left[ \left\Vert \frac C N \sum_{i=1}^N \epsilon_i \int_{0}^{t_i} k(\tau, \cdot) d\tau \right\Vert_{\rkhs}^2 \right] }
		\tag{Jensen inequality}
		\\
		&=
		\sqrt{ \E_{\epsilon_i} \left[ \frac{C^2}{N^2} \sum_{i,j=1}^N \epsilon_i \epsilon_j
			\left\langle
			\int_{0}^{t_i} k(\tau, \cdot) d\tau, \:
			\int_{0}^{t_j} k(\tau, \cdot) d\tau
			\right\rangle_{\rkhs}
			\right] }
		\\
		&=
		\sqrt{ \frac{C^2}{N^2} \sum_{i}^N
			\left\Vert
			\int_{0}^{t_i} k(\tau, \cdot) d\tau
			\right\Vert_{\rkhs}^2
		}
		\tag{$\epsilon_i$ are uncorrelated}
		\\
		&\leq
		\sqrt{ \frac{C^2}{N^2} \sum_{i}^N
			\left(
			\int_{0}^{t_i} \left\Vert k(\tau, \cdot) \right\Vert_{\rkhs} d\tau
			\right)^2
		}
		\tag{triangle inequality}
		\\
		&\leq
		\sqrt{ \frac{C^2}{N^2} \sum_{i}^N
			\left(
			\int_{0}^{T} \sup_{t \in [0, T]} \left\Vert k(t, \cdot) \right\Vert_{\rkhs} d\tau
			\right)^2
		}
		\\
		&=
		TC N^{-1/2} \sup_{t \in [0, T]} \left\Vert k(t, \cdot) \right\Vert_{\rkhs}.
	\end{align*}
	Recognizing that $k(t, \cdot)$ is a bounded-everywhere reproducing kernel completes the proof.
\end{proof}

\begin{lemma}
	\label{lemma:rkhs_rademacher_vector}
	Define $\rkhs_{M'}$ as in \cref{lemma:min_norm_existance}.
	For any $0 < C < \infty$,
	define the sets
	\[
	\begin{gathered}
		\sF^C_{M'} := \left\{ (\dot\vx, \dot\vmu, \dot\vy) \in \rkhs_{M'} \: : \: \Vert (\dot\vx, \dot\vmu, \dot\vy) \Vert_{\rkhs_D} \leq C \right\}
		\\
		\tint\sF^C_{M'} := \left\{ \tint_{0}^{(\cdot)} (\dot\vx(\tau), \dot\vmu(\tau), \dot\vy(\tau)) d\tau \: : \: (\dot\vx, \dot\vmu, \dot\vy) \in \sF \right\}
	\end{gathered}
	\]
	Then $\widehat\rademacher_N(\sF^C_{M'}) \lesssim C M N^{-1/2}$ and $\widehat\rademacher_N(\tint\sF^C_{M'}) \lesssim T M C N^{-1/2}$.
\end{lemma}
\begin{proof}
	The proof follows a standard summation argument for Rademacher complexity:
	\begin{align*}
		&\widehat\rademacher(\sF^C_{M'})
		\\
		:=&
		\E \left[
		\sup_{(\dot\vx, \dot\vmu, \dot\vy) \in \rkhs_{M'}} \frac 1 N \sum_{i=1}^N \left(
		\sum_{j_x=1}^{\Mx} \epsilon_{ij_x} \dot\evx^{(j_x)}(t_i) +
		\sum_{j_y=1}^{\Mx} \epsilon_{ij_y} \dot\evy^{(j_y)}(t_i) +
		\sum_{j_z=1}^{\Mx} \epsilon_{ij_z} \dot\evz^{(j_z)}(t_i)
		\right) \right]
		\tag{$\epsilon_{ij_x}, \epsilon_{ij_y}, \epsilon_{ij_z} \simiid \mathrm{Rad}$}
		\\
		\leq&
		\sum_{j_x=1}^{\Mx} \E \left[ \sup_{\dot\evx^{(j_x)} \in \rkhs} \frac 1 N \sum_{i=1}^N \epsilon_{ij_x} \dot\evx^{(j_x)}(t_i) \right]
		+
		\sum_{j_y=1}^{\Mx} \E \left[ \sup_{\dot\evy^{(j_y)} \in \rkhs} \frac 1 N \sum_{i=1}^N \epsilon_{ij_y} \dot\evy^{(j_y)}(t_i) \right]
		\\ &+
		\sum_{j_z=1}^{\My} \E \left[ \sup_{\dot\evz^{(j_z)} \in \rkhs} \frac 1 N \sum_{i=1}^N \epsilon_{ij_z} \dot\evz^{(j_z)}(t_i) \right]
		\\
		\lesssim& \: C M N^{-1/2},
		\tag{\cref{lemma:rkhs_rademacher}}
	\end{align*}
	where the last inequality comes from the fact that $\Vert (\dot\vx, \dot\vmu, \dot\vy) \Vert_{\rkhs_{M'}} \leq C$
	implies that
	$\Vert \dot\vx^{(m)} \Vert_{\rkhs}$,
	$\Vert \dot\vmu^{(m)} \Vert_{\rkhs}$,
	$\Vert \dot\vy^{(m)} \Vert_{\rkhs} \leq C$
	for all $m$.
	An analogous proof holds for $\widehat\rademacher(\tint\sF^C_{M'})$.
\end{proof}

\begin{lemma}
	\label{lemma:rademacher_loss}
	Denote $\rkhs_{M'}$ as in \cref{lemma:set_s_closed}.
	For any $(\dot\vx, \dot\vmu, \dot\vy) \in \sobolev^{2,2}_M$,
	$\hat\vmu_0 \in \R^{\Mx}$, $\hat\vy_0 \in \R^{\My}$,
	define the differential equation error function
	\begin{equation}
		\begin{gathered}
			e_{\dot\vx, \dot\vmu, \dot\vy, \hat\vmu_0, \hat\vy_0}(\cdot) :=
			\left\Vert \begin{bmatrix}
				\dot\vx(\cdot) - \mF\left( \vx, \vmu, \vy\right) \\
				\dot\vmu(\cdot) - \vGh\left( \vx, \vmu, \vy\right) \\
				\mH\left( \vx, \vmu, \vy\right)
			\end{bmatrix} \right\Vert_\infty
			\\
			\vx(t) := \vx_0 + \int_0^t \dot\vx(\tau) d\tau,
			\quad
			\vmu(t) := \hat\vmu_0 + \int_0^t \dot\vmu(\tau) d\tau,
			\quad
			\vy(t) := \hat\vy_0 + \int_0^t \dot\vy(\tau) d\tau,
		\end{gathered}
		\label{eqn:diffeq_err_fn}
	\end{equation}
	For any $0 < C < \infty$,
	define $\sG^{C,K}$ as the set of error functions
	\[
	\left\{
	e_{\dot\vx, \dot\vmu, \dot\vy, \hat\vmu_0, \hat\vy_0}(\cdot) \: : \:
	\Vert (\dot\vx, \dot\vmu, \dot\vy)  \Vert_{\sobolev^{2,2}_M} \leq C,
	\Vert \hat\vmu_0 \Vert_\infty \leq K,
	\Vert \hat\vy_0 \Vert_\infty \leq K
	\right\}.
	\]
	Then every error function in $\sG^{C,K}$ is bounded by some constant $\tilde C$
	and $\widehat\rademacher(\sG^{C,K}) \lesssim C N^{-1/2}$.
\end{lemma}
\begin{proof}
	Note that each error function in $\sG^{C,K}$ is a Lipschitz function
	($\Vert \cdot \Vert_\infty$),
	each of which is applied to the summation of two sub-functions:
	\begin{enumerate}
		\item a $(\dot\vx(\cdot), \dot\vmu(\cdot), \dot\vy(\cdot)) \in \rkhs_{M'}$ with norm less than $C$
		(i.e. an element of $\sF^C_{M'}$, as defined in \cref{lemma:rkhs_rademacher_vector}), and
		\item lipschitz functions ($\mF$, $\vGh$, $\mH$) applied to the integral of a vector-valued RKHS function ($\dot\vx, \dot\vmu, \dot\vy$) with norm less than $C$
		(i.e. a Lipschitz function applied to an element of $\tint\sF^C_{M'}$, as defined in \cref{lemma:rkhs_rademacher_vector}).
	\end{enumerate}
	Boundedness of the error functions falls from the fact $(\vx, \vmu, \vy)$ are bounded, $\hat\vmu_0, \hat\vy_0$ are bounded, and $\mF$, $\vGh$, $\mH$ are continuous (and thus bounded over $[0, T]$).
	The Rademacher complexity falls from standard Lipschitz and summation rules:
	$
	\widehat\rademacher(\sG^{C,K}) \lesssim \widehat\rademacher(\sF^C_{M'}) + \widehat\rademacher(\tint\sF^C_{M'}) \lesssim C M N^{-1/2}.
	$
\end{proof}

Now we are ready to prove \cref{thm:consistency}.

\begin{proof}[Proof of \cref{thm:consistency}]
	We begin by noting that \[ \begin{aligned}
		B :=& \inf_{(\vx, \vmu, \vy \in \sS}
		\tsum_{m=1}^{\Mx}\, \Vert \evxh^{(m)} \Vert^2_{\rkhs}
		+ \tsum_{m=1}^{\Mx}\, \Vert \evmuh^{(m)} \Vert^2_{\rkhs}
		+ \tsum_{m=1}^{\My}\, \Vert \evyh^{(m)} \Vert^2_{\rkhs}
		\\
		=& \inf_{(\vx, \vmu, \vy \in \sS} \Vert (\dot\vx, \dot\vmu, \dot\vy) \Vert_{\rkhs_{M'}}
		< \infty
	\end{aligned} \]
	is implied by \cref{lemma:sobolev_solutions}.
	Let $(\vx^*, \vmu^*, \vy^*)$ be some element in $\sS$ that achieves this infimum
	(the existence of which is guaranteed by \cref{lemma:min_norm_existance}).
	We know that
	$\Vert (\dot\vxh_N, \dot\vmuh_N, \dot\vyh_N) \Vert_{\rkhs_{M'}} \leq
	\Vert (\dot\vx^*, \dot\vmu^*, \dot\vy^*) \Vert_{\rkhs_{M'}} = B$---%
	since $(\vx^*, \vmu^*, \vy^*)$ satisfies the constraints for \cref{eq:erm-norm}---%
	and thus $(\dot\vxh_N, \dot\vmuh_N, \dot\vyh_N) \in \sF^B_{M'}$ (as defined by \cref{lemma:rkhs_rademacher_vector}).

	Defining $e_{\dot\vxh_N, \dot\vmuh_N, \dot\vyh_N, \hat\vmu_0, \hat\vy_0}(\cdot)$ as in \cref{lemma:rademacher_loss},
	we have that $e_{\dot\vxh_N, \dot\vmuh_N, \dot\vyh_N, \hat\vmu_0, \hat\vy_0}(t_i) = 0$
	for each $t_i$ in $\train$.
	Applying a standard uniform large law argument \citep[e.g.][Thm.~4.2]{wainwright2019high}
	we have that, for any $\delta > 0$,
	\begin{align*}
		\int_0^T e_{\dot\vxh_N, \dot\vmuh_N, \dot\vyh_N, \hat\vmu_0, \hat\vy_0}(\tau) d\tau
		&=
		\left\vert
		\frac{1}{N} \sum_{i=1}^N e_{\dot\vxh_N, \dot\vmuh_N, \dot\vyh_N, \hat\vmu_0, \hat\vy_0}(t_i)
		-
		\int_0^T e_{\dot\vxh_N, \dot\vmuh_N, \dot\vyh_N, \hat\vmu_0, \hat\vy_0}(\tau) d\tau
		\right\vert
		\\
		&\leq
		2\E_{t_i} \left[ \widehat\rademacher(\sG^{B,K}) \right] + \delta
	\end{align*}
	with probability $1 - 2\exp(-\frac{N\delta^2}{8 \tilde C^2})$
	(where $\tilde C$ is the constant defined in \cref{lemma:rademacher_loss}).
	Since $\widehat\rademacher(\sG^{B,K}) \lesssim B M N^{-1/2}$ (\cref{lemma:rademacher_loss}),
	we have that
	$\int_0^T e_{\dot\vxh_N, \dot\vmuh_N, \dot\vyh_N, \hat\vmu_0, \hat\vy_0}(\tau) d\tau
	\overset{\mathrm{a.s.}}{\longrightarrow} 0$,
	which implies that $\lim_{N\to\infty} (\vxh_N, \vmuh_N, \vyh_N)$
	satisfies the differential equation almost everywhere.

	Define $(\dot\vx, \dot\vmu, \dot\vy)$ as the continuously-differentiable representative of the function
	$\lim_{N\to\infty} (\dot\vxh_N, \dot\vmuh_N, \dot\vyh_N)$
	(\citep[see e.g.][Thm.~8.2]{brezis2011functional}).
	Since $(\dot\vx, \dot\vmu, \dot\vy)$ is continuously-differentiable and satisfies the differential equation everywhere,
	it must also be an element of $\sS$.
	All together, this implies that $\Vert (\dot\vx, \dot\vmu, \dot\vy) \Vert_{\rkhs_{M'}} \geq B$,
	and so
	$\Vert (\dot\vx, \dot\vmu, \dot\vy) \Vert_{\rkhs_{M'}} =
	\Vert \lim_{N\to\infty} (\dot\vxh_N, \dot\vmuh_N, \dot\vyh_N) \Vert_{\rkhs_{M'}} =
	B$.
\end{proof}

	
	\newpage
	\bibliographystyle{qe}
	\bibliography{references}

	\newpage
	\begin{center}
		{\LARGE Supplement to “Solving Models of Economic Dynamics with Ridgeless Kernel Regressions”}
	\end{center}
	\vspace{1em}
	\begin{center}
	Mahdi Ebrahimi Kahou\textsuperscript{1}, 
	Jesse Perla\textsuperscript{2}, 
	Geoff Pleiss\textsuperscript{3,4}
	
	\vspace{0.5em}
	
	\textsuperscript{1}Department of Economics, Bowdoin College. \\
	\textsuperscript{2}Vancouver School of Economics, University of British Columbia. \\
	\textsuperscript{3}Department of Statistics, University of British Columbia. \\
	\textsuperscript{4}Vector Institute.
	\end{center}
	\noindent
	This supplement contains the details of the robustness checks on our algorithm and provides additional applications alongside those presented in the paper.

	\setcounter{section}{0} 
	\renewcommand{\thesection}{\arabic{section}} 
	\newpage

	\section{Robustness}\label{sec:app-variations}
This section provides robustness checks and an exploration of sample efficiency for the neoclassical growth model.
\subsection{Sparse training data and data efficiency}\label{sec:sparse}
\begin{figure}[H]
	\centering
	\includegraphics[width=\textwidth]{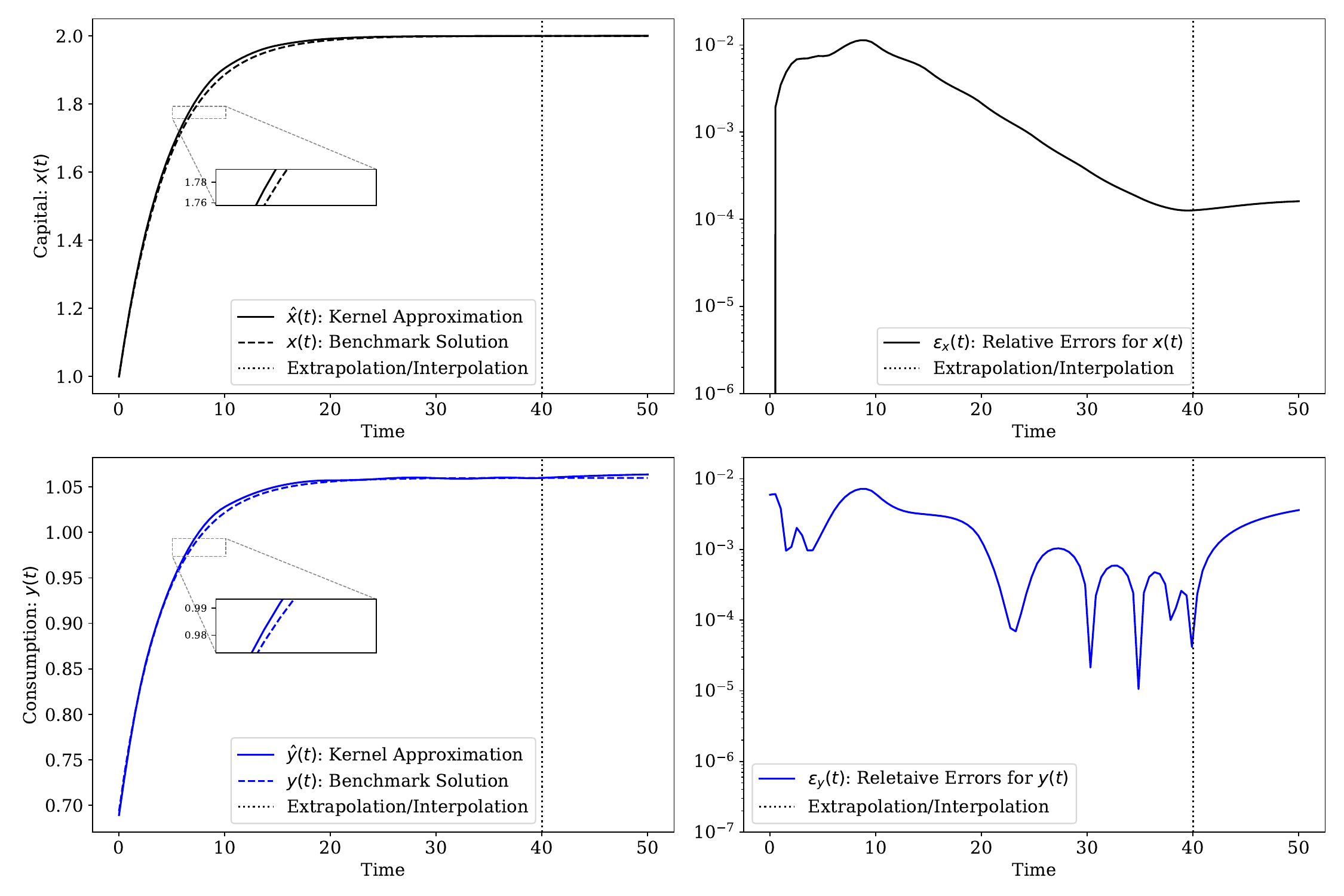} 
	\caption{Solution of the neoclassical growth model (\cref{eq:ngm-f1,eq:ngm-f2,eq:ngm-f3} of the main paper) without imposing the transversality condition (\cref{eq:ngm-f4} of the main paper), for sparse training data $
		\mathcal{D} := \{0, 1, 3, 5, 10, 15, 20, 25, 30, 35, 38, 40\}.$ The top panels show the results for capital, $x(t)$ and associated relative errors, and the bottom two show the results for consumption, $y(t)$. Accurate solutions can be obtained even with a very sparse $\mathcal{D}$.} 
	\label{fig:neoclassical_growth_model_sparse}
\end{figure}

Since the number of parameters in the optimization grows linearly with the number of grid points, the cardinality of the training set, $\mathcal{D}$, becomes an impediment in higher dimensions with many state and jump variables. Therefore, obtaining accurate approximate solutions with sparse training data is crucial.

\cref{fig:neoclassical_growth_model_sparse} shows the result of the neoclassical growth model (i.e., \cref{eq:ngm-f1,eq:ngm-f2,eq:ngm-f3} of the main paper) solved with  
$\mathcal{D} := \{0, 1, 3, 5, 10, 15, 20, 25, 30, 35, 38, 40\}$.
The top-left panel shows the approximate and benchmark capital paths, denoted by $\hat{x}(t)$ and $x(t)$, respectively. The top-right panel shows the relative errors between the approximate and benchmark solutions for capital, denoted by $\varepsilon_x(t)$. The bottom-left panel shows the approximate and benchmark consumption paths, denoted by $\hat{y}(t)$ and $y(t)$, respectively. The bottom-right panel shows the relative errors between the approximate and benchmark solutions for consumption, denoted by $\varepsilon_y(t)$.

These results show that one can obtain very accurate approximate solution, even with a very sparse training data.  
\subsection{Robustness to the choice of the kernel and kernel parameters} \label{sec:app-choice-kernel}
\cref{tab:robustness} shows the result of the approximate solution of the neoclassical growth model, described in \cref{eq:ngm-f1,eq:ngm-f2,eq:ngm-f3} of the main paper, for different Mat\'{e}rn kernels and kernel parameters. 

The first three rows report the performance of the approximate solutions using three different kernels. We present the maximum and minimum absolute values of the relative errors for both the capital path, $\hat{x}(t)$, and the consumption path, $\hat{y}(t)$. The first row shows the baseline solution using the Mat\'{e}rn kernel with $\nu = \frac{1}{2}$. The second and third rows present results for the Mat\'{e}rn kernels with $\nu = \frac{3}{2}$ and $\nu = \frac{5}{2}$, respectively.
\renewcommand{\arraystretch}{1.2}
\begin{table}[H]
	\begin{center}
		\resizebox{\textwidth}{!}{\begin{tabular}{c}
				\begin{tabular}{lrrrrr}
\toprule
$\nu$ & $\ell$ & Max of Rel. Error: $\hat{x}(t)$ & Max of Rel. Error: $\hat{y}(t)$ & Min of Rel. Error: $\hat{x}(t)$ & Min of Rel. Error: $\hat{y}(t)$ \\
\midrule
$1/2$ & 10 & 1.8e-03 & 2.9e-03 & 7.1e-05 & 9.6e-07 \\
$3/2$ & 10 & 5.9e-04 & 3.0e-02 & 1.5e-05 & 3.4e-06 \\
$5/2$ & 10 & 1.4e-04 & 2.4e-02 & 2.7e-05 & 6.0e-08 \\
$1/2$ & 2 & 3.1e-03 & 2.8e-03 & 1.3e-07 & 5.5e-07 \\
$1/2$ & 20 & 1.9e-03 & 8.2e-02 & 7.6e-05 & 2.9e-05 \\
\bottomrule
\end{tabular}

		\end{tabular}}
	\end{center}
	\hspace*{2mm}
	\caption{The robustness of the approximate solutions of the neoclassical growth model (i.e., \cref{eq:ngm-f1,eq:ngm-f2,eq:ngm-f3} of the main paper) is tested using different  Mat\'{e}rn kernels, $\nu = \frac{1}{2}, \frac{3}{2}, \frac{5}{2}$, and length scales $\ell = 2, 10, 20$. 
	}
	\label{tab:robustness}
\end{table}
\renewcommand{\arraystretch}{1.0}
The last two rows show the performance of the approximate solutions for two different {\it length scales}, $\ell = 2$, and $\ell = 20$.  

Throughout these experiments, we achieve highly accurate approximate solutions. Therefore, the results demonstrate insensitivity to the selection of  Mat\'{e}rn kernels and the length scales.
\subsection{Smaller time horizons: accurate short-run dynamics}\label{sec:app-far-steady-state}
One might suspect that achieving an accurate optimal solution, which does not violate the transversality condition, is only possible if one uses a large time horizon in the training data. For instance, we use $\train = \{0,1,2,\ldots,30\}$ to obtain the results depicted in \cref{fig:ngm-results}. In this experiment, we establish that we can still achieve accurate short-run dynamics by using a smaller time horizon.
\begin{figure}[H]
	\centering
	\includegraphics[width=\textwidth]{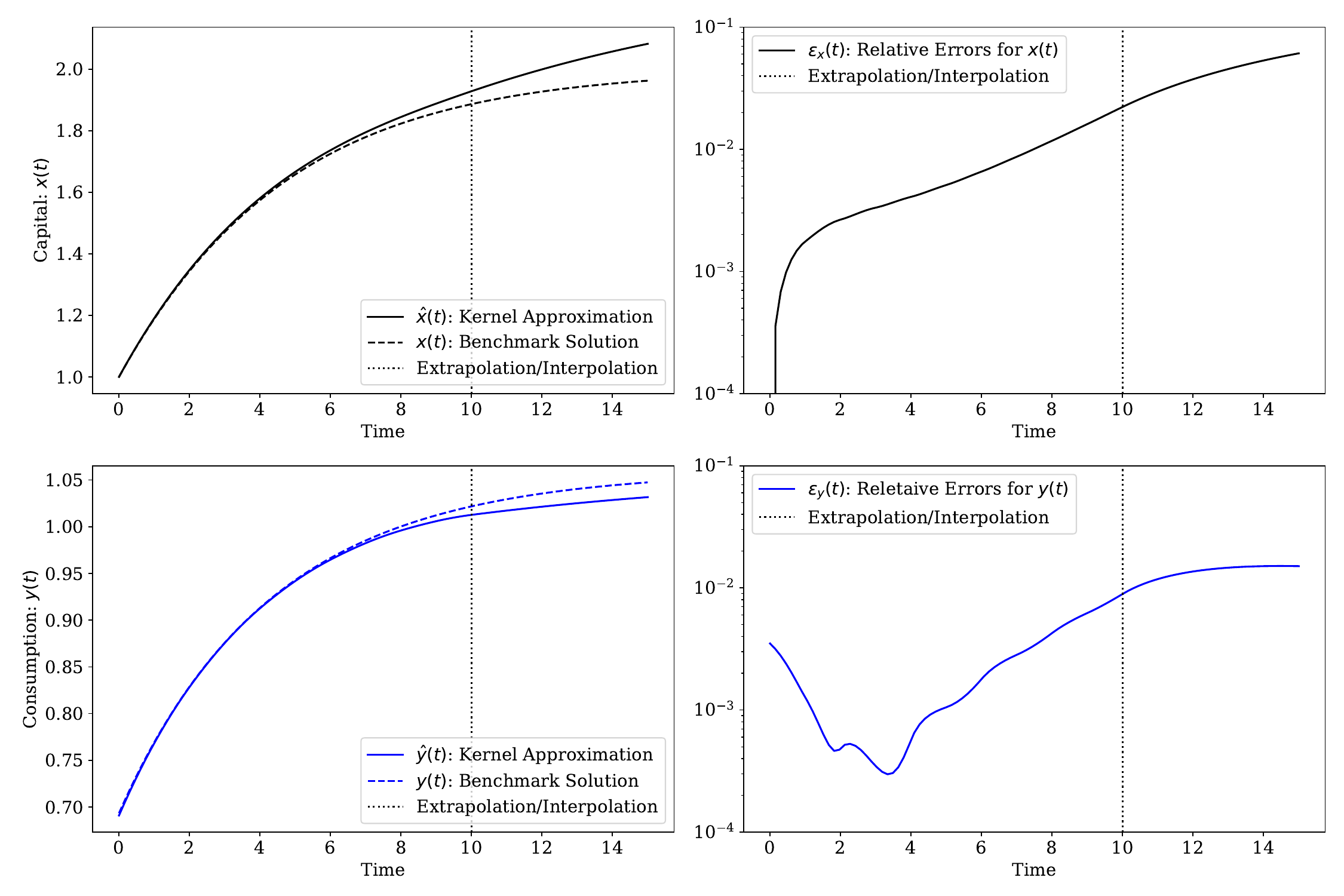} 
	\caption{ Solution of the neoclassical growth model (\cref{eq:ngm-f1,eq:ngm-f2,eq:ngm-f3} of the main paper) without imposing the transversality condition (\cref{eq:ngm-f4} of the main paper), for training data with a smaller time horizon $
		\mathcal{D} := \{0,\cdots,10\}.$ The top panels show the results for capital, $x(t)$ and associated relative errors, and the bottom two show the results for consumption, $y(t)$. Accurate short-run dynamics obtained even a smaller time horizon. 
	}
	\label{fig:neoclassical_growth_model_far_steady_state}
\end{figure}

\cref{fig:neoclassical_growth_model_far_steady_state} shows the approximate solutions for the neoclassical growth model (i.e., \cref{eq:ngm-f1,eq:ngm-f2,eq:ngm-f3} of the main paper) for training data with smaller time horizon, defined as $\mathcal{D}:=\{0,1,2,\cdots,10\}$. The top-left panel shows the approximate and benchmark capital paths, denoted by $\hat{x}(t)$ and $x(t)$, respectively. The top-right panel shows the relative errors between the approximate and benchmark solutions for capital, denoted by $\varepsilon_x(t)$. The bottom-left panel shows the approximate and benchmark consumption paths, denoted by $\hat{y}(t)$ and $y(t)$, respectively. The bottom-right panel shows the relative errors between the approximate and benchmark solutions for consumption, denoted by $\varepsilon_y(t)$.

	
\numberwithin{equation}{section} 
\renewcommand{\theequation}{S\arabic{equation}}
\section{More Applications}
In this section, we discuss additional applications. In \cref{sec:human-capital}, we solve a model that incorporates the time evolution of human capital and its interaction with physical capital in economic growth. The model includes seven variables: two state variables along with their co-state variables, and three jump variables. In \cref{sec:optimal-advertising}, we solve an optimal advertising model based on the work of \citet{sethi1973optimal}.

\subsection{Human capital and growth} \label{sec:human-capital}
In this example, we solve the neoclassical growth model with human and physical capital as illustrated in \cite{acemoglu2008introduction}. The optimal paths for the state variables $\vx(t) := \left[x_k(t),x_h(t)\right]$, co-state variables $\vmu(t) := \left[\mu_k(t),\mu_h(t)\right]$, and jump variables $\vy(t) := \left[y_c(t), y_{k}(t), y_{h}(t)\right]$ solve 

\begin{align}
	\dot{x}_k(t) &= y_k(t) - \delta_k x_k(t),\label{eq:human-capital-x-k-dot}\\
	\dot{x}_h(t) &= y_h(t) - \delta_h x_h(t),\label{eq:human-capital-x-h-dot}\\
	\dot{\mu}_k(t) &= r \mu_k(t) - \mu_k(t) \bigl[ f_1(x_k(t),\,x_h(t)) - \delta_k \bigr],\label{eq:human-capital-mu-k-dot}\\
	\dot{\mu}_h(t) &= r \mu_h(t) - \mu_h(t) \bigl[ f_2(x_k(t),\,x_h(t)) - \delta_h \bigr],\label{eq:human-capital-mu-h-dot}\\
	0 &= \mu_k(t) y_c(t) - 1,\label{eq:human-capital-shadow-price}\\
	0 &= \mu_k(t) - \mu_h(t),\label{eq:human-capital-shadow-prices-equality}\\
	0 &= f(x_k(t),\,x_h(t)) - y_c(t) - y_k(t) - y_h(t),\label{eq:human-capital-feasibility-condition}
\intertext{with two transversality conditions}
	0 &= \lim_{t\rightarrow \infty} e^{-rt} x_k(t) \mu_k(t),\label{eq:human-capital-transversality-k}\\
	0 &= \lim_{t\rightarrow \infty} e^{-rt} x_h(t) \mu_h(t),\label{eq:human-capital-transversality-h}
\end{align}

for given initial conditions $x_k(0) = x_{k_0}$, $x_h(0) = x_{h_0}$.

The production function is defined as $f\left(x_k(t),x_h(t)\right) = x_k(t)^{a_k}x_h(t)^{a_h}$. Here,  $f_1\left(\cdot,\cdot\right)$ is the derivative with respect to the first input and $f_2\left(\cdot,\cdot\right)$ is the derivative with respect to the second input. The two constants in the production function, $a_k$ and $a_h$, are positive numbers, such that $a_k+a_h <1$. Additionally, $\delta_k>0$, $\delta_h>0$, and $r>0$.

Human capital is denoted by $x_h(t)$, physical capital by $x_k(t)$, consumption by $y_c(t)$, investment in human capital by $y_{h}(t)$, and investment in physical capital by $y_{k}(t)$. Here $\mu_k(t)$ and $\mu_{h}(t)$ are the co-state variables.

This problem is more challenging than the neoclassical growth model introduced in \cref{sec:results} of the main paper because it has twice the number of state and co-state variables, and adds additional algebraic equations, \cref{eq:human-capital-shadow-prices-equality,eq:human-capital-feasibility-condition}.  Furthermore, the dynamics become coupled through no-arbitrage conditions between investment in physical and human capital.

For an arbitrary initial condition, this formulation leads to a time-zero discontinuity where $x_k(0+\epsilon)$ and $x_h(0+\epsilon)$ jump to the solution manifold.  We choose $x_k(0)$ solve for a $x_h(0)$ consistent the no-arbitrage $f_h(k(\epsilon), h(\epsilon))  - \delta_h =  f_k(k(\epsilon), h(\epsilon)) -  \delta_k$.  The dynamics show that even in cases with more challenging coupling, kernel methods can find consistent solutions without imposing the transversality conditions.

In this experiment we, use $\delta_k = 0.1$, $\delta_h = 0.05$, $\alpha_k = \frac{1}{3}$,  $\alpha_h = \frac{1}{4}$, $r = 0.11$, $x_{k_0} = 1.5$, and $x_{h_0} = 1.37 $ as the numerical values for the economic parameters. We use $\mathcal{D}=\{0,1,\cdots,80\}$ as the training data.  In order to stabilize the solution given the coupling of the differential equations, we also found it necessary to add to the objective $\lambda_p \times (||i_k||_{\rkhs} + ||i_h||_{\rkhs}+ ||c||_{\rkhs})$ with $\lambda_p = 5\times10^{-3} $. 

\begin{figure}[H]
	\centering
	\includegraphics[width=\textwidth]{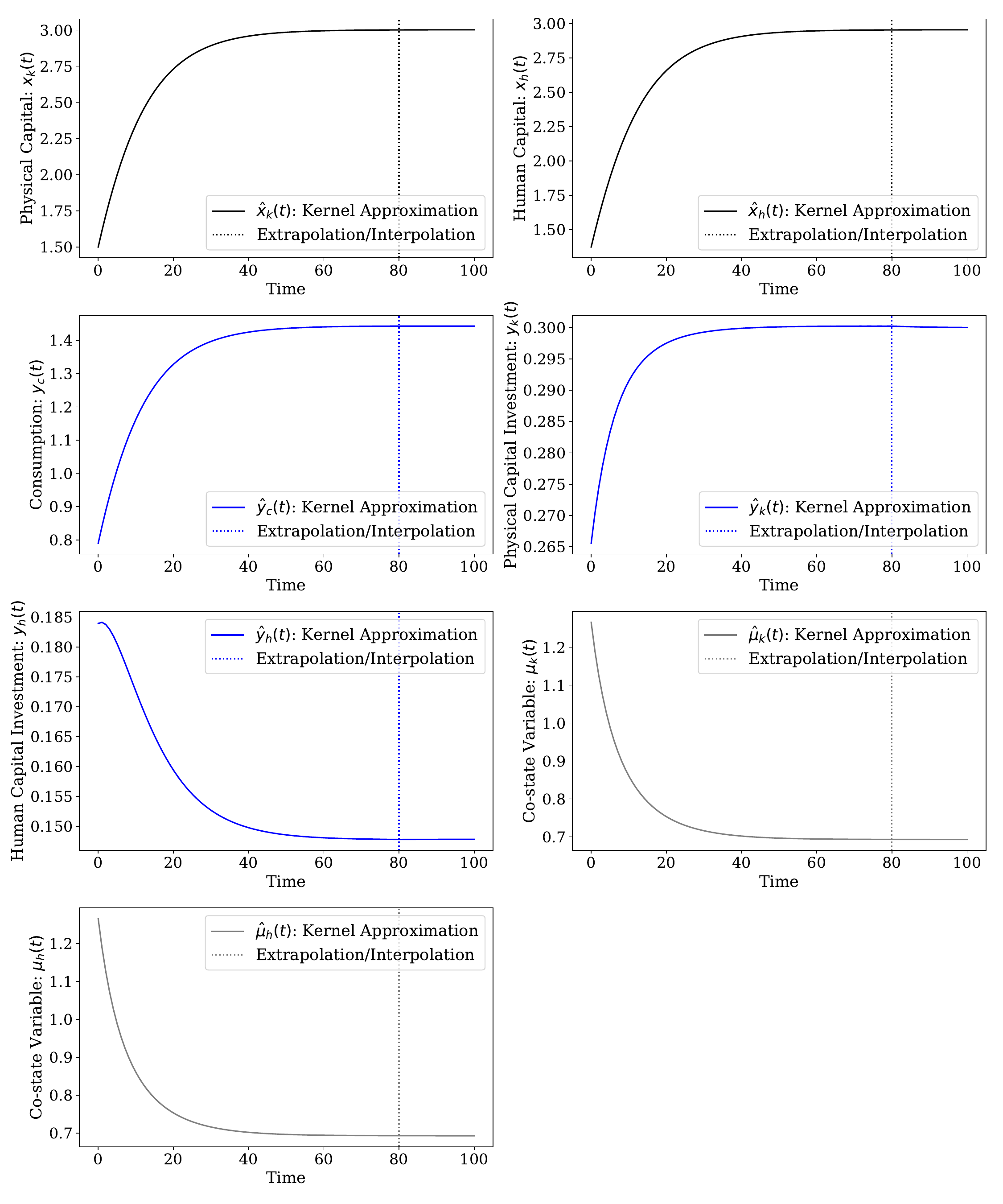} 
	\vspace{-7mm}
	\caption{Solution of the growth model with human capital (see \cref{eq:human-capital-x-k-dot,eq:human-capital-x-h-dot,eq:human-capital-mu-k-dot,eq:human-capital-mu-h-dot,eq:human-capital-shadow-price,eq:human-capital-shadow-prices-equality,eq:human-capital-feasibility-condition})  without imposing the transversality conditions. A non-explosive approximate solution is obtained; as shown, all paths converge to the correct steady states.
	}
	\label{fig:human_capital}
\end{figure}

\cref{fig:human_capital} shows the approximate paths of physical capital ($\hat{x}_k(t)$), human capital ($\hat{x}_h(t)$), consumption ($\hat{y}_c(t)$), investment in physical capital ($\hat{y}_k(t)$), and investment in human capital ($\hat{y}_h(t)$), along with the co-state variables ($\hat{\mu}_k(t)$ and $\hat{\mu}_h(t)$). The vertical dashed line shows the boundary between the interpolation and extrapolation regions.

{\it The correct set of steady states.~} How do we know the approximate solutions converge to the correct set of steady states? A set of solutions that violates the transversality conditions is characterized by paths such that $\displaystyle\lim_{t\rightarrow \infty} e^{-r t}\mu_k(t)x_k(t) \neq 0$ and $\displaystyle\lim_{t \rightarrow \infty} e^{-r t}\mu_h(t)x_h(t) \neq 0$. Since the approximate solutions shown in \cref{fig:human_capital} do not exhibit this behavior, we can be confident that they converge to the correct set of steady states.

\subsection{Optimal Advertising}
\label{sec:optimal-advertising}
In this example we solve an optimal advertising model based on the classical model of expenditure on advertising introduced in \citep{vidale1957operations}. The optimal paths $\vx(t)$, $\vy(t)$, and $\vmu(t)$, solve
\begin{align}
	\dot{x}(t) & = \left[1-x(t)\right]y(t)-\beta x(t)
	\label{eq:optima-advertising-x-dot},\\
	\dot{\mu}(t) &= r\mu(t) -\gamma + \beta\mu(t)+ \mu(t)y(t)
	\label{eq:optima-advertising-mu-dot}\\
	0 & = y(t)^{\frac{1-\kappa}{\kappa}}- \kappa \mu(t)\left[1-x(t)\right]\label{eq:optima-shadow-price}
\end{align}
for a given initial condition $\vx(0)= \vx_0$, and a transversality condition
\begin{align}
	\lim_{t\rightarrow \infty} e^{-r t}x(t) \mu(t) = 0,\label{eq:optima-advertising-TVC}
\end{align}
$x$ represents the market share of the company, $y$ is a variable corresponding to advertising expenditure, and $\mu$ is the co-state variable. 

The parameter $\kappa$ is a constant between $0$ and $1$, $\beta$ is strictly positive, $r$ is the discount rate, the constant $\gamma$ is defined as $\gamma := \frac{\beta+r}{c}$, and $c$ is the cost of advertising. See \citep{weber2006infinite,sethi1973optimal} for a detailed treatment of this problem.

\begin{figure}[H]
	\centering
	\includegraphics[width=\textwidth]{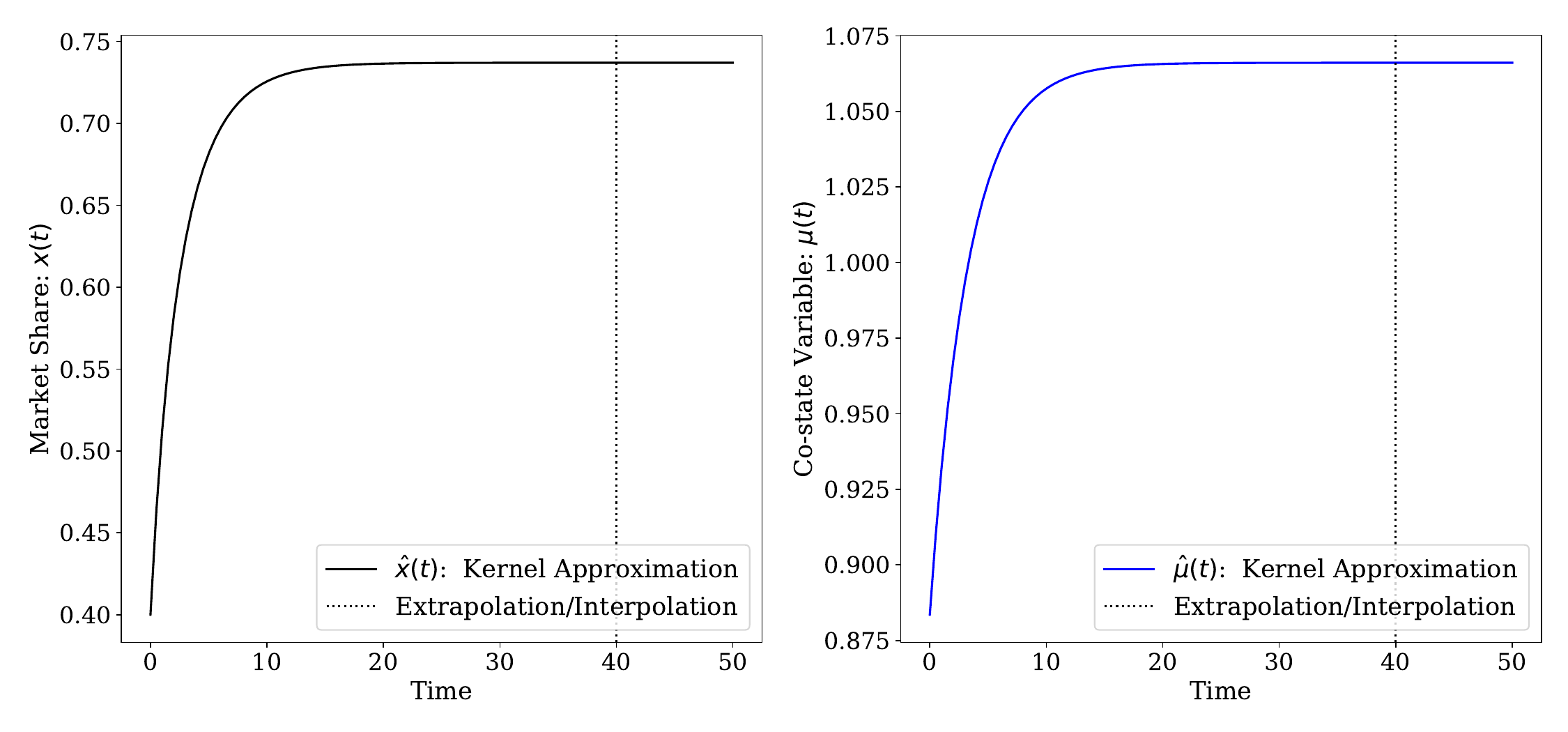} 
	\vspace{-7mm}
	\caption{Solution of the optimal advertising model (i.e., \cref{eq:optima-advertising-x-dot,eq:optima-advertising-mu-dot,eq:optima-shadow-price}) without imposing the transversality condition.}		
	\label{fig:optimal_advertising}
\end{figure}
In this example we use $x_0 = 0.4$, $r = 0.11$, $c = 0.5$, $\beta = 0.05$, $\kappa = 0.5$, and $\gamma = \frac{\beta+r}{c} = 0.32$ as the numerical values for the parameters. We use $\mathcal{D} = \{0,1,\cdots,40\}$ as the training data.

{Results.~}
\cref{fig:optimal_advertising} shows the approximate market share, denoted by $\hat{x}(t)$, and the approximate co-state variable, denoted by $\hat{\mu}(t)$, obtained using a Mat\'ern kernel with $\nu = \frac{1}{2}$. The vertical dashed line indicates the boundary between the interpolation and extrapolation regions. Despite not applying the transversality condition (\cref{eq:optima-advertising-TVC}), the kernel approximation accurately recovers the optimal solution.

{\it The correct set of steady states.~} How do we know the approximate solution recovers the optimal solution? As shown in \cref{fig:optimal_advertising}, the approximate state, and co-state variable approach a finite number. Therefore,  $\lim_{t\rightarrow \infty} e^{-r t}\hat{\mu}(t) \hat{x}(t) = 0$.  


\end{document}